\PassOptionsToPackage{hyphens}{url} 
\documentclass[%
 aip,
 amsmath,amssymb,
 reprint,%
]{revtex4-2}
\usepackage{graphicx}
\usepackage{dcolumn}
\usepackage{bm}
\usepackage[utf8]{inputenc}
\usepackage[T1]{fontenc}
\usepackage{mathptmx,comment}
\usepackage{amsmath, amssymb, mathtools, bm, etoolbox, comment}
\usepackage{setspace}
\usepackage{mathrsfs}
\usepackage{stmaryrd}
\usepackage{amsthm}
\usepackage{amsfonts}
\usepackage{graphicx}
\usepackage[colorinlistoftodos]{todonotes}
\usepackage{tikz-cd}   
\usepackage{booktabs}       
\usepackage{nicefrac}       
\usepackage{microtype}      
\usepackage[utf8]{inputenc} 
\usepackage[T1]{fontenc}    
\usepackage{hyperref}       
\usepackage{url}            
\usepackage{lingmacros}
\usepackage{tree-dvips} 
\usepackage{nccmath}
\usepackage{bbm}

\DeclareMathOperator*{\loggrad}{log-grad}

\usepackage[utf8]{inputenc} 
\usepackage[noend]{algpseudocode}
\usepackage{algorithm}
\usepackage{caption}
\usepackage{subcaption}
\usepackage{color}
\usepackage{enumitem}

\usepackage{amsmath}

\begin{document}

\preprint{AIP/123-QED} 
\title[COVID-19 second wave mortality in Europe and the United States]{COVID-19 second wave mortality in Europe and the United States}

\author{Nick James}
\affiliation{ 
School of Mathematics and Statistics, University of Sydney, NSW, 2006, Australia}%
\author{Max Menzies}
\email{max.menzies@alumni.harvard.edu}
\affiliation{%
Yau Mathematical Sciences Center, Tsinghua University, Beijing, 100084, China}%
\author{Peter Radchenko}
\affiliation{School of Business, University of Sydney, NSW, 2006, Australia}

\date{24 December 2020}
\begin{abstract}

This paper introduces new methods to analyze the changing progression of COVID-19 cases to deaths in different waves of the pandemic. First, an algorithmic approach partitions each country or state's COVID-19 time series into a first wave and subsequent period. Next, offsets between case and death time series are learned for each country via a normalized inner product. Combining these with additional calculations, we can determine which countries have most substantially reduced the mortality rate of COVID-19. Finally, our paper identifies similarities in the trajectories of cases and deaths for European countries and U.S. states. Our analysis refines the popular conception that the mortality rate has greatly decreased throughout Europe during its second wave of COVID-19; instead, we demonstrate substantial heterogeneity throughout Europe and the U.S. The Netherlands exhibited the largest reduction of mortality, a factor of 16, followed by Denmark, France, Belgium, and other Western European countries, greater than both Eastern European countries and U.S. states. Some structural similarity is observed between Europe and the United States, in which Northeastern states have been the most successful in the country. Such analysis may help European countries learn from each other's experiences and differing successes to develop the best policies to combat COVID-19 as a collective unit.

\end{abstract}

\maketitle

\begin{quotation}

Europe is experiencing a substantial second wave of COVID-19.  Epidemiologists have attributed this to loosening of both government restrictions and individual precautions. \cite{Looi2020} This reflects the ongoing struggle to balance the spread of the virus and allowing public life to return to normal, a year into the pandemic. \cite{bbcnormal} The mortality rate of the disease, and how it changes over time, plays a critical role in this debate. Identifying countries that have reduced their mortality rate in subsequent waves of the disease is therefore of great relevance to policymakers. In the public conception, health experts and journalists alike have noted a substantial decrease in the mortality of COVID-19 in Europe's second wave. \cite{FTsecondsurge} This paper aims to refine this conception with a mathematical analysis on a country-by-country basis. Instead, we show a considerable variance in the reduction of mortality in different countries' second waves. Most wealthy countries in Europe, with the notable exceptions of Germany and Sweden, have drastically reduced their mortality rate during their second wave. Less wealthy European countries, as well as many U.S. states, have seen a less notable reduction.

\end{quotation}

\section{Introduction}
\label{sec:intro}

Government responses to COVID-19 have varied substantially, both from country to country and as time has progressed. Early responses included banning travel \cite{bbccloseborders_2020} and establishing test and trace programs, \cite{guardian_2020} followed by lockdowns as cases rose, often implemented too late. \cite{nyt2020,Scally2020} Due to the economic consequences and unpopularity of lockdowns, the U.S. states prioritized reopening well before the virus was entirely suppressed. \cite{wapo_allreopen} Such responses to the virus, highly varying with time, have created \emph{first and subsequent waves} of the outbreak in most countries, with subsequent waves often exhibiting greater case numbers than the first. \cite{atlantic_secondsurge, james2020covidusa}

Fortunately, subsequent waves of COVID-19 have featured a reduced mortality rate in many countries. \cite{Griffin2020} Explanations for this include the development of new treatments for COVID-19 as time passes \cite{Remdesivir,Bloch2020,toczilizumab,Cao2020} and under-reporting of true case numbers in the first wave. \cite{underreporting} We believe this is the first paper to perform a mathematical analysis to quantify the reduction of mortality on a country-by-country basis throughout Europe, and compare European countries with U.S. states. We demonstrate significant heterogeneity in Europe, with wealthier Western European countries having reduced their mortality rate more drastically than less wealthy European countries or U.S. states.

To perform our analysis, we use new and recently introduced techniques in \emph{time series analysis}. Time series analysis has been frequently used in epidemiology, \cite{Hethcote2000, Chowell2016} including to study the Zika virus, \cite{Biswas2020, Morrison2020} Ebola, \cite{Funk2018, Mhlanga2019} and COVID-19. \cite{Manchein2020, Machado2020, chaos1James2020, Blasius2020,Perc2020} Nonlinear dynamics researchers apply a wide range of methods, including power-law models, \cite{Blasius2020, Manchein2020, Vazquez2006} distance analysis, \cite{james2020covidusa, Moeckel1997, Szkely2007, Mendes2018, Mendes2019,James2020_nsm} forecasting models,\cite{Perc2020} and network models. \cite{Shang2020, Karaivanov2020} In this work, we apply the algorithmic framework introduced in Ref. \onlinecite{james2020covidusa} to partition COVID-19 case time series into a first wave and a subsequent period (the latter could consist of a single second wave or multiple waves). Later, we consider all the European countries and U.S. states in conjunction, and identify similarities in their case and death trajectories via \emph{clustering}. We implement \emph{hierarchical clustering}; this technique has been used in a wide variety of epidemiological applications. \cite{Madore2007, Kretzschmar2009, Alashwal2019, Muradi2015, Rizzi2010, Machado2020}

This paper is structured as follows: in Section \ref{sec:offset}, we study the reduction in mortality rate between the first and subsequent waves of COVID-19 among European countries and U.S. states. This relies on a new framework for partitioning time series and learning appropriate offsets. In Section \ref{sec:Trajectories}, we study all European countries and U.S. states in conjunction, clustering case and death trajectories to elucidate similarities across the two groups. We summarize our findings regarding COVID-19 mortality and trajectories across first and subsequent waves in Section \ref{sec:Conclusion}.

\section{First and subsequent wave analysis}
\label{sec:offset}

In this section, we describe a mathematical framework to analyze the changing mortality relative to first and subsequent waves of COVID-19. We first apply our constructions on an individual country-by-country basis and then collectively compare all the European countries and U.S. states. Our list of European countries comes from the United Nations, \cite{Europelist} except we exclude Russia and the Holy See. We consider all U.S. states plus the District of Columbia (D.C.). This gives us 93 total countries and states.

\subsection{Methodology: determination of offsets and mortality ratios}
\label{sec:secondsurgemethod}

Let $x(t), y(t)$ be the new daily case and death time series, respectively, of a single country or state, $t=0,...,T$. In this paper, data for every country and state spans 01/21/2020 to 11/25/2020, a period of 310 days. The end date corresponds to the last week that the European Centre for Disease Prevention and Control (ECDC) provided daily data updates. \cite{datachanged} That is, $T=309$ for every time series.

First, we apply the methodology developed in Ref. \onlinecite{james2020covidusa} to divide each country into first and subsequent waves of the disease. Specifically, we apply a smoothing filter to the case time series $x(t)$ followed by a two-step algorithm to output an alternating sequence of local maxima (peaks) and minima (troughs), beginning with a trough at $t=0$, where there are zero cases. Further details are provided in Appendix \ref{appendix:turningpoint}. We apply this only to the case counts, as the death counts are much sparser.  Just two European countries and three U.S. states in our analysis are assigned a sequence that consists of just one trough at $t=0$, and one peak. These countries and states are determined to still be in their first wave of COVID-19. For every other country, we have at least one non-trivial trough. Let $T_1$ be the first non-trivial trough, or the second trough after $t=0$. This marks the end of the complete first wave in the corresponding country. We refer to the period $t=0,...,T_1$ as the first wave, and $t=T_1+1,...,T$ as the subsequent period. In particular, we consider any second and third wave as one period, to be compared with the distinguished first wave.

We aim to analyze the changing mortality rate by comparing the first wave and the subsequent period. Indeed, we wish to understand if countries were able to learn and adapt their treatment of the disease after the end of the first wave. To appropriately compare the case and death time series, we must calculate an offset in time between cases and deaths for each country. For this purpose, we use \emph{normalized inner products}, and can either assume that there is a single offset for the entire time window, or two offsets - one each for the first wave and subsequent period.

For each country, let $\tau$ be the optimal single offset between the case and death time series. We define this as the optimal value of the normalized inner product
\begin{align}
<x(0:T-\tau),y(\tau:T)>_n =\\
\frac{x(0)y(\tau)+...+x(T-\tau)y(T)}{(x(0)^2+...+x(T-\tau)^2)^\frac{1}{2}(y(\tau)^2+...+y(T)^2)^\frac{1}{2}}.
\end{align}
We have chosen these normalized inner products to have maximal value 1 if and only if there is a proportionality relation $y(t)=kx(t+\tau)$ for all $t=0,...,T-\tau$ for some constant $k>0$. Indeed, we are seeking the offset in time where deaths are most closely proportional to cases. This is more suitable than other metrics, such as correlation or distance correlation.  \cite{Szkely2007} Correlation or distance correlation would each return maximal value 1 if $y=kx + b$ for an additional constant $b$, which is unsuitable.

Having determined this offset, we can define
\begin{align}
M_1&= \dfrac{\sum_{t=\tau}^{T_1+\tau}  y(t)}{\sum_{t=0}^{T_1} x(t)},\\
M_2&=\frac{\sum_{t=T_1+1+\tau}^{T} y(t)}{\sum_{t=T_1+1}^{T-\tau} x(t)}.
\end{align}
These record the mortality rates for each country over the first wave and subsequent period, respectively, taking into account a single offset between case and death counts.

Alternatively, we can determine a pair of two offsets $\lambda_1$ and $\lambda_2$. The first offset $\lambda_1$ is chosen to maximize the normalized inner product 
\begin{align}
<x(0:T_1),y(\lambda_1:T_1+\lambda_1)>_n.    
\end{align}
while the second offset $\lambda_2$ is chosen to maximize the normalized inner product 
\begin{align}
<x(T_1+1:T - \lambda_2),y(T_1+\lambda_2:T)>_n. 
\end{align}
With these two offsets, we can define 
\begin{align}
N_1&= \dfrac{\sum_{t=\lambda_1}^{T_1+\lambda_1}  y(t)}{\sum_{t=0}^{T_1} x(t)},\\
N_2&=\frac{\sum_{t=T_1+1+\lambda_2}^{T} y(t)}{\sum_{t=T_1+1}^{T-\lambda_2} x(t)}.
\end{align}
These record the mortality rates for each country over the first wave and subsequent period, respectively, taking into account two respective offsets between case and death counts. Each country or state is assigned its own value of $T_1, \tau, M_1, M_2, \lambda_1, \lambda_2, N_1, N_2$, while $T=309$ is fixed for all of them. We refer to $\tau, \lambda_1, \lambda_2$ as \emph{offsets} and $M_1/M_2$ and $N_1/N_2$ as  \emph{one- and two-offset mortality ratios}, respectively.

Finally, we can generate convenient scatter plots to show the case and death counts of the first wave and the subsequent period. Using the two offsets $\lambda_1, \lambda_2$, we can simply plot the set of first wave values $\{(x(t),y(t+\lambda_1)) \in \mathbb{R}^2: t=0,...,T_1\}$ and subsequent period values $\{(x(t),y(t+\lambda_2))\in \mathbb{R}^2: t=T_1,...,T - \lambda_2\}$. In Section \ref{secondsurgeresults}, we plot these two periods in red and blue, respectively, and include their centroid (that is, average) and convex hull. \cite{convexhull}

\subsection{Offsets and mortality ratio results}
\label{secondsurgeresults}

\begin{figure*}
    \centering
    \begin{subfigure}[b]{0.3\textwidth}
        \includegraphics[width=\textwidth]{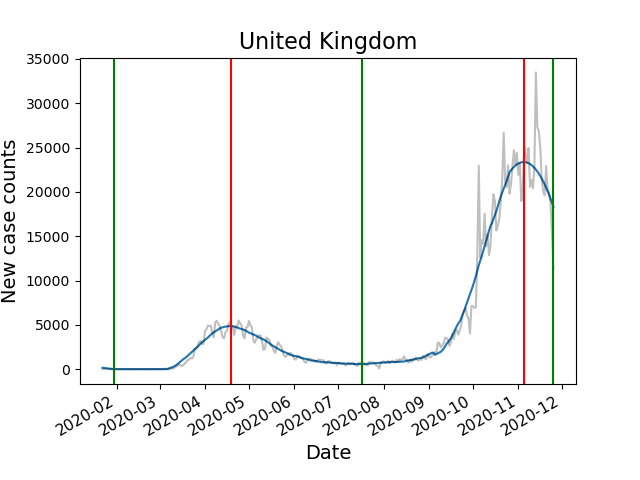}
        \caption{ }
        \label{fig:UK_cases}
    \end{subfigure}
    \begin{subfigure}[b]{0.3\textwidth}
        \includegraphics[width=\textwidth]{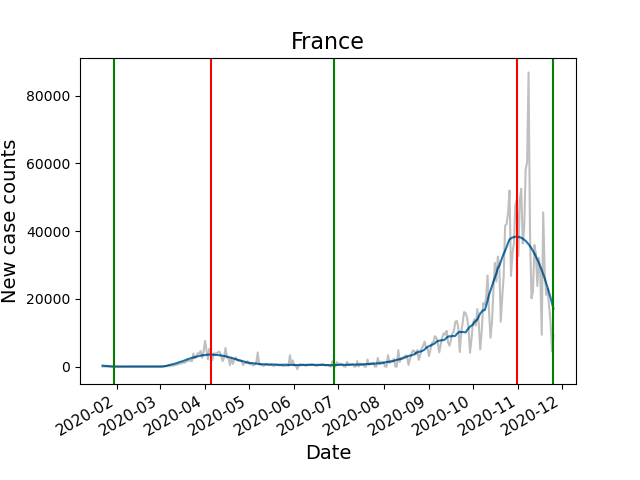}
        \caption{}
        \label{fig:France_cases}
    \end{subfigure}
    \begin{subfigure}[b]{0.3\textwidth}
        \includegraphics[width=\textwidth]{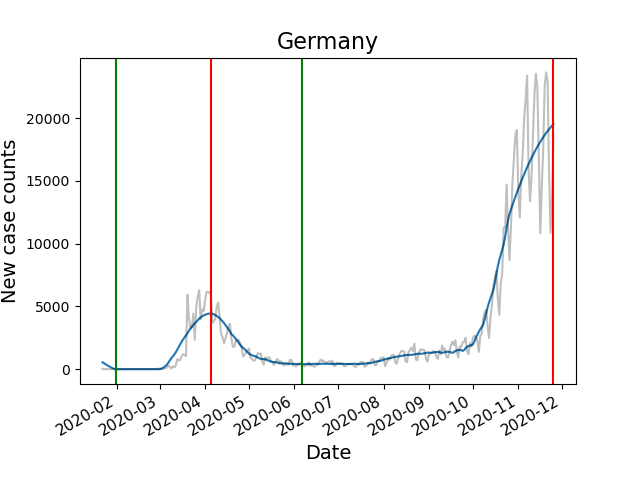}
        \caption{}
        \label{fig:Germany_cases}
    \end{subfigure}    
    \begin{subfigure}[b]{0.3\textwidth}
        \includegraphics[width=\textwidth]{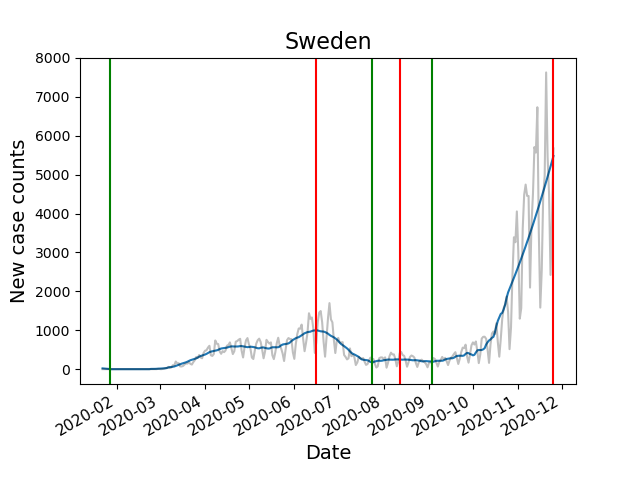}
        \caption{ }
        \label{fig:Sweden_cases}
    \end{subfigure}
    \begin{subfigure}[b]{0.3\textwidth}
        \includegraphics[width=\textwidth]{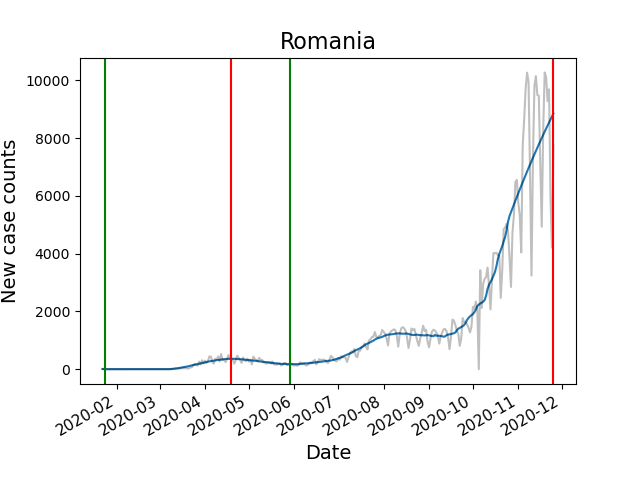}
        \caption{}
        \label{fig:Romania_cases}
    \end{subfigure}
    \begin{subfigure}[b]{0.3\textwidth}
        \includegraphics[width=\textwidth]{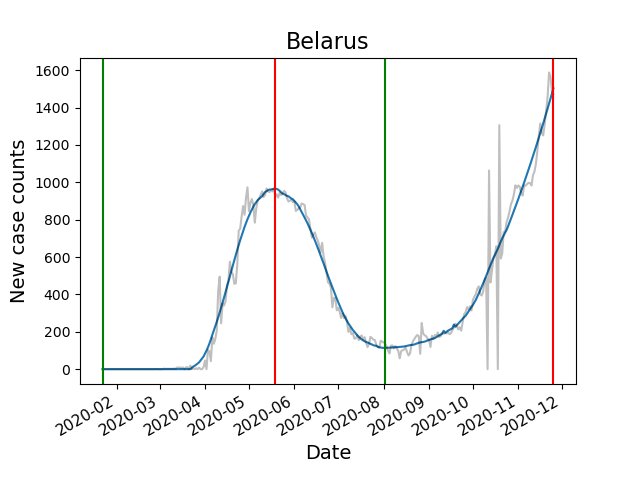}
        \caption{}
        \label{fig:Belarus_cases}
    \end{subfigure}
    \caption{Smoothed case time series and identified turning points for various European countries: (a) the United Kingdom (b) France (c) Germany (d) Sweden (e) Romania (f) Belarus. Green and red vertical lines denote algorithmically detected troughs and peaks, respectively. The peaks and troughs partition the year into different waves of the disease on a country-by-country basis. Each aforementioned country experienced more than one wave, with greater case numbers observed in subsequent waves than the first.}
    \label{fig:Cases_time_series}
    \end{figure*}

\begin{figure*}
    \centering
    \begin{subfigure}[b]{0.3\textwidth}
        \includegraphics[width=\textwidth]{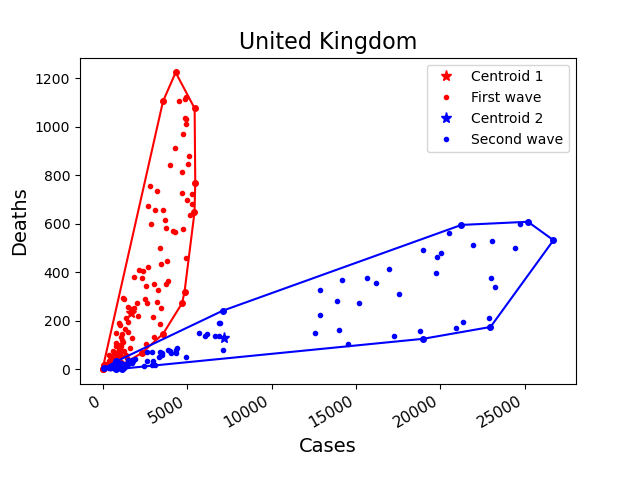}
        \caption{ }
        \label{fig:UK_convex_hull}
    \end{subfigure}
    \begin{subfigure}[b]{0.3\textwidth}
        \includegraphics[width=\textwidth]{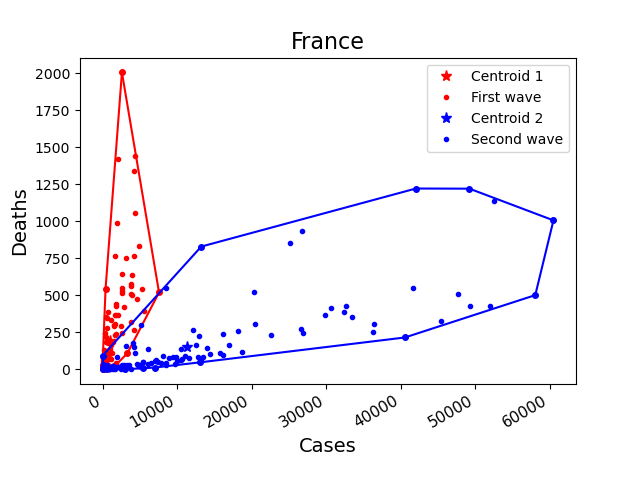}
        \caption{}
        \label{fig:France_convex_hull}
    \end{subfigure}
    \begin{subfigure}[b]{0.3\textwidth}
        \includegraphics[width=\textwidth]{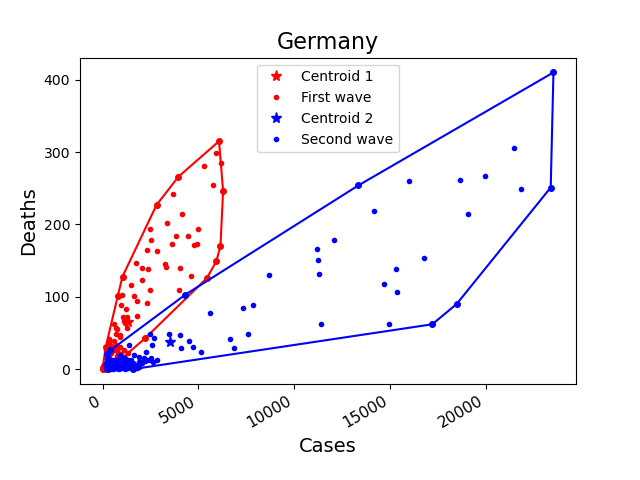}
        \caption{}
        \label{fig:Germany_convex_hull}
    \end{subfigure}    
        \begin{subfigure}[b]{0.3\textwidth}
        \includegraphics[width=\textwidth]{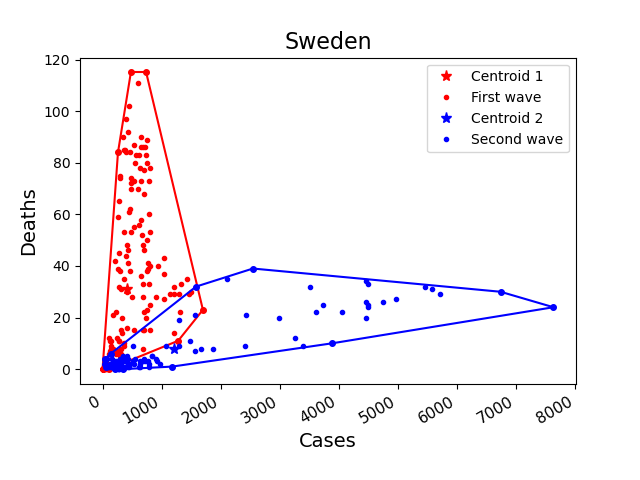}
        \caption{ }
        \label{fig:Sweden_convex_hull}
    \end{subfigure}
    \begin{subfigure}[b]{0.3\textwidth}
        \includegraphics[width=\textwidth]{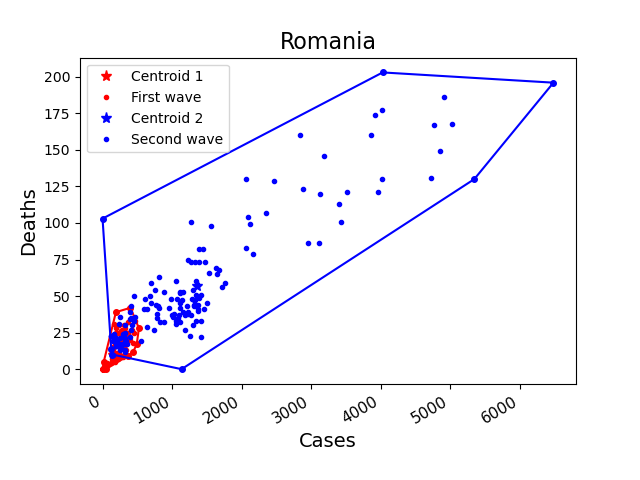}
        \caption{}
        \label{fig:Romania_convex_hull}
    \end{subfigure}
    \begin{subfigure}[b]{0.3\textwidth}
        \includegraphics[width=\textwidth]{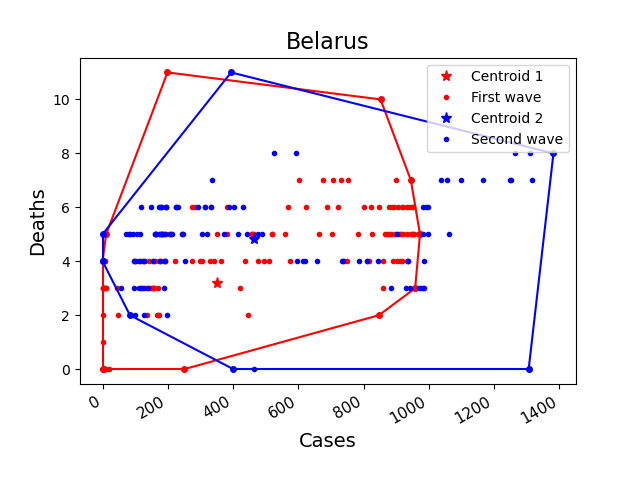}
        \caption{}
        \label{fig:Belarus_convex_hull}
    \end{subfigure}
    \caption{Scatter plots of cases and deaths for the first wave and subsequent period, for (a) the United Kingdom (b) France (c) Germany (d) Sweden (e) Romania (f) Belarus. The first wave data is plotted in red, the subsequent period in blue, with two offsets taken into account. The plots include centroids and convex hulls for the two time periods. A steeper slope of the convex hull represents a higher mortality rate. The wealthier countries display a greater progression from cases to deaths in the first wave than the subsequent period. For Belarus, the progression is largely the same.}
    \label{fig:Convex_hull_plots}
\end{figure*}

Figure \ref{fig:Cases_time_series} displays new case time series and algorithmically determined turning points (peaks and troughs) for six European countries. The first non-trivial trough $T_1$ splits each time series up into its first wave and subsequent period. All these countries display a similar structure - they experience a first wave in cases followed by at least one more significant subsequent wave. The United Kingdom (U.K.), France and Germany, displayed in Figures \ref{fig:UK_cases},  \ref{fig:France_cases}, and \ref{fig:Germany_cases}, respectively, highlight a characteristic case trajectory for wealthier European countries, in which there are two predominant waves in COVID-19 cases. Less developed European countries such as Romania and Belarus, displayed in Figures  \ref{fig:Romania_cases} and \ref{fig:Belarus_cases}, respectively, produce a similar new case trajectory to the wealthier European countries, yet many with a smaller first wave, like Romania.

In Tables \ref{tab:Europe_offsets} and \ref{tab:USA_offsets}, respectively, we record the values of $\tau$, $M_1/M_2$, $\lambda_1, \lambda_2$ and $N_1/N_2$ for European countries and U.S. states, respectively. First, we notice significant variance between European countries in the one-offset mortality ratios $M_1/M_2$. The Netherlands has the highest ratio of 16.2, followed by Denmark, France, Belgium, Spain, the U.K., Ireland, Andorra, Finland and Norway, all of which are wealthy countries in Western or Northern Europe. These countries are indicated to have drastically reduced COVID-19 mortality between their first wave and subsequent period. By contrast, the smallest mortality ratios are exhibited by Belarus, Malta, Iceland, Latvia, Albania and Serbia, most of which are less developed. Notable wealthy countries with comparatively low mortality ratios include Germany (3.6) and Sweden (3.8). These countries have not reduced their mortality ratio as much relative to their first wave.

We obtain broadly similar results in the two-offset mortality ratios $N_1/N_2$. Once again, the Netherlands has the highest ratio, followed by France, Sweden, Belgium, Finland, Ireland, Norway, Spain, Andorra, Denmark and the U.K. Sweden has changed its position drastically due to significant differences in the optimized values of $\tau, \lambda_1, \lambda_2$. Malta, Belarus, Iceland, Latvia, Albania and Serbia have the lowest two-offset mortality ratios. Broadly, but not universally, the two methods give similar results, indicating the robustness of the methodology.

Such heterogeneity is also observed for the United States. The U.S. states with the greatest one-offset mortality ratio are Vermont (9.2), New Jersey, New York and Connecticut, all Northeastern states, with similar results observed for the two-offset mortality ratio. No U.S. state reduced its mortality as much as the Netherlands, Denmark, France, Belgium, Spain, the U.K. or Ireland.

We can further elucidate the differences between first and subsequent wave mortality by examining the case-death scatter plots in Figure  \ref{fig:Convex_hull_plots}. As described in Section \ref{sec:secondsurgemethod}, we plot both first and subsequent waves accounting for two offsets, and include the centroid and convex hull of each set, understood as a subset of $\mathbb{R}^2$. The U.K., France, Germany and Sweden, displayed in Figures \ref{fig:UK_convex_hull}, \ref{fig:France_convex_hull}, \ref{fig:Germany_convex_hull} and \ref{fig:Sweden_convex_hull}, respectively, show a similar pattern. These countries experience a significantly higher mortality rate during their first wave than subsequent period. This finding is consistent among many of the more developed, wealthier, European countries, though the separation of the scatter plot regions is sharper for the U.K. and France than for Germany and Sweden, consistent with their mortality ratios. The pattern among several less developed European countries such as Romania and Belarus is markedly different, as seen in Figures  \ref{fig:Romania_convex_hull} and \ref{fig:Belarus_convex_hull}. Romania is one of several eastern European countries with a smaller first wave, while Belarus exhibits no real change in the progression from cases to deaths at all.

\begin{figure*}
    \centering
    \begin{subfigure}[b]{0.49\textwidth}
        \includegraphics[width=\textwidth]{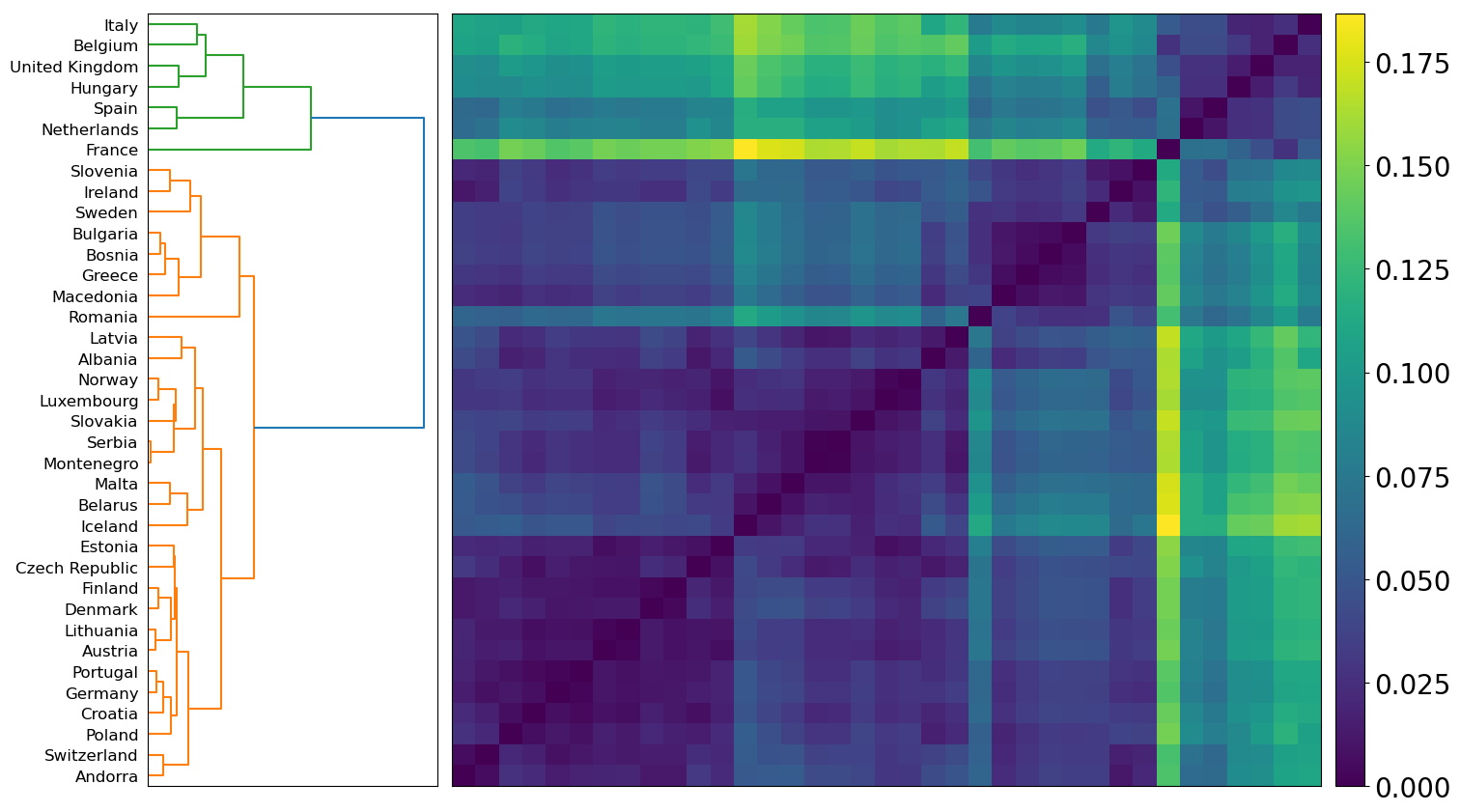}
        \caption{}
        \label{fig:Europe_M1M2_1}
    \end{subfigure}
    \begin{subfigure}[b]{0.49\textwidth}
        \includegraphics[width=\textwidth]{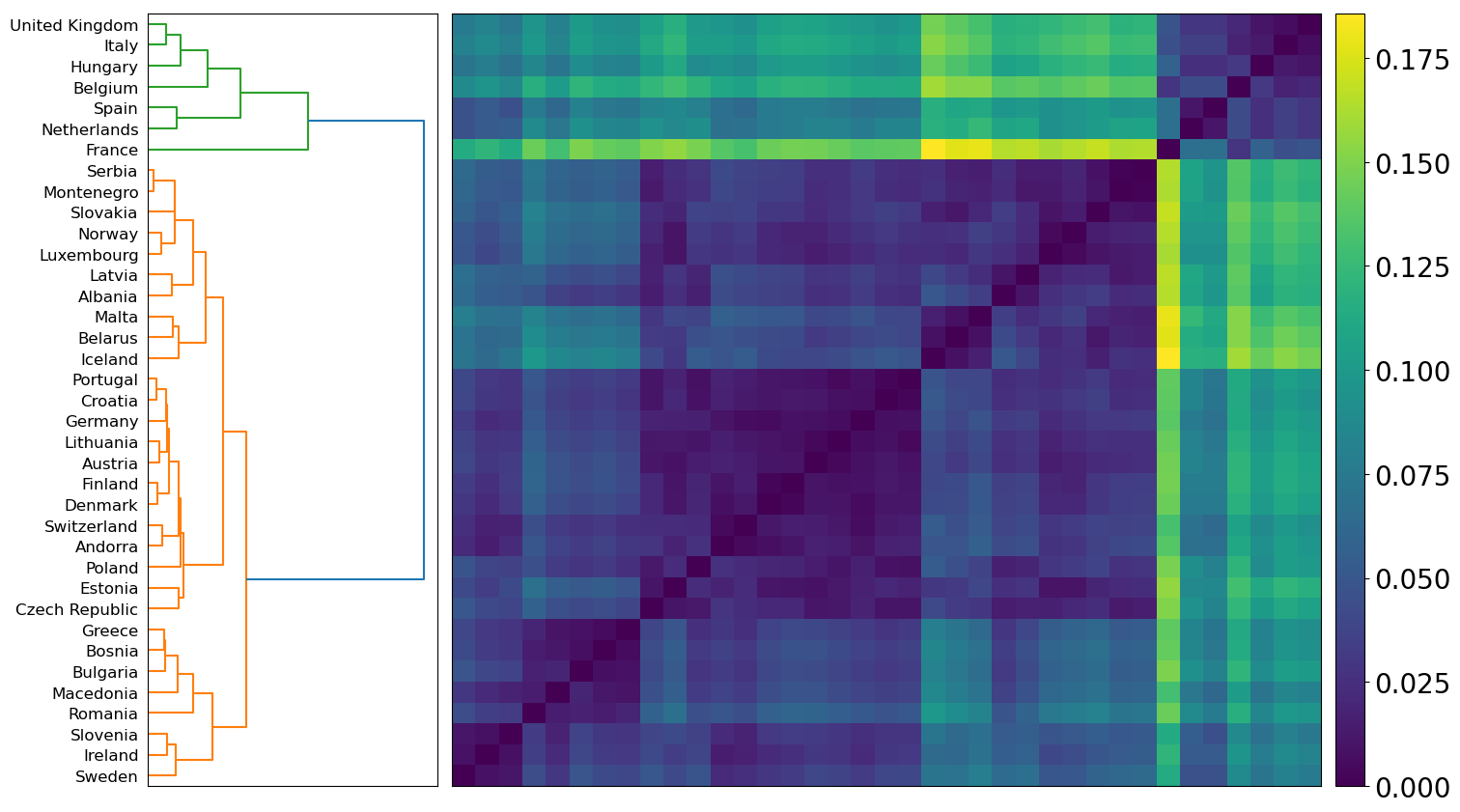}
        \caption{}
        \label{fig:Europe_M1M2_2}
    \end{subfigure}    
    \begin{subfigure}[b]{0.49\textwidth}
        \includegraphics[width=\textwidth]{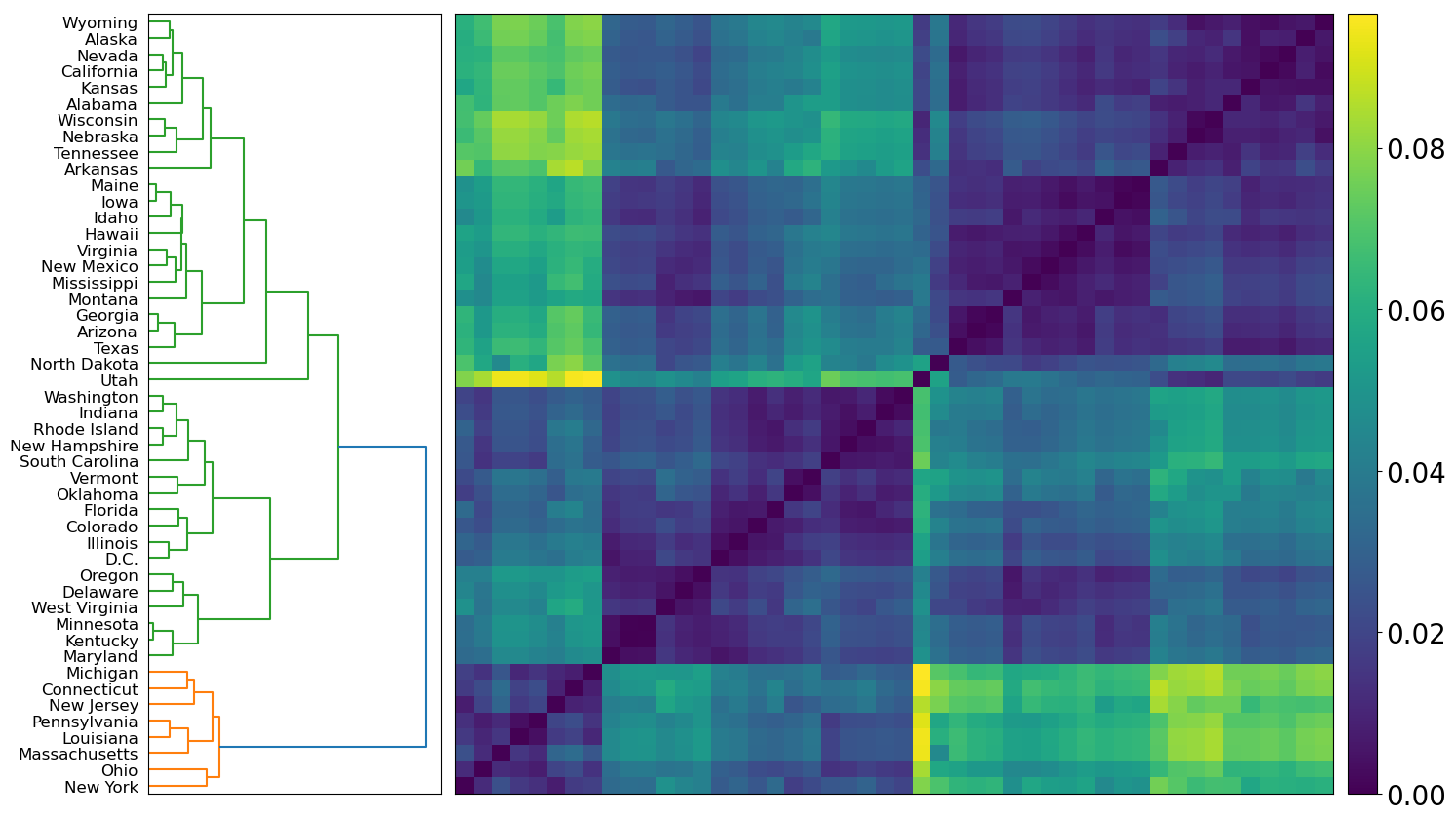}
        \caption{ }
        \label{fig:USA_M1M2_1}
    \end{subfigure}
    \begin{subfigure}[b]{0.49\textwidth}
        \includegraphics[width=\textwidth]{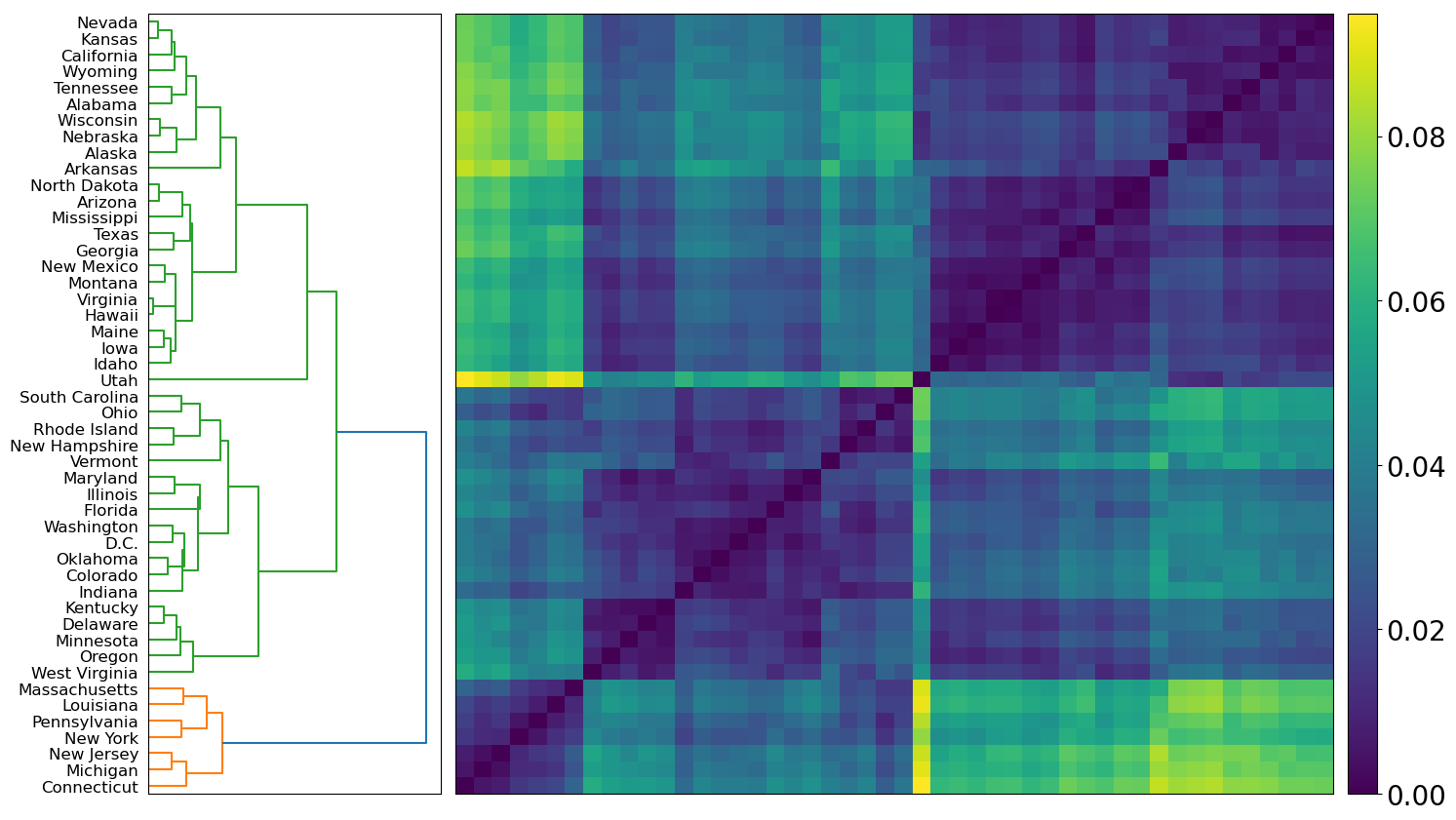}
        \caption{ }
        \label{fig:USA_M1M2_2}
    \end{subfigure}
    \caption{Hierarchical clustering on the one-offset and two-offset mortality pairs $(M_1,M_2)$ and $(N_1,N_2)$. European countries are clustered relative to $(M_1,M_2)$ in (a) and $(N_1,N_2)$ in (b). U.S. states are clustered relative to $(M_1,M_2)$ in (c) and $(N_1,N_2)$ in (d). Moldova, Ukraine, Missouri, North Carolina and South Dakota do not have a subsequent period and so are excluded. A similar structure is observed between European countries and U.S. states, where Western European countries and Northeastern U.S. states display more significant reductions in their mortality rates.}
    \label{fig:Dendrogram_M1M2}
\end{figure*}

We extend our analysis of mortality by implementing hierarchical clustering on the ordered pairs $(M_1,M_2)$ and $(N_1,N_2)$, understood as elements of $\mathbb{R}^2$. We retain more information this way without taking the quotients. Figures \ref{fig:Europe_M1M2_1} and \ref{fig:Europe_M1M2_2} demonstrate  a consistent structure among European countries across both our one offset and two offset models. Both figures contain two clusters; a smaller cluster and a larger cluster with two sub-clusters. In each dendrogram, the smaller cluster contains the same members, Belgium, France, Hungary, Italy, the Netherlands, Spain and the U.K. Excluding Hungary, these countries have much in common. They are all wealthy western European countries that have among the highest $M_1/M_2$ and $N_1/N_2$ ratios. They experienced severe first waves in both cases and deaths and implemented harsh lockdown procedures in response. Even Hungary has comparatively high mortality ratios relative to the rest of Europe.

Figures \ref{fig:USA_M1M2_1} and \ref{fig:USA_M1M2_2} display the cluster structure among U.S. states, with striking similarities to the European countries. Once again, there are two clusters in each figure, one larger cluster with two subclusters and one smaller cluster. Again, the smaller cluster highlights a collection of anomalous states, predominantly located in the Northeastern United States. Connecticut, Massachusetts, New Jersey, New York, and Pennsylvania feature in both smaller clusters. Like western Europe, these states experienced early and severe first waves. There is a consistent theme among European countries and  U.S. states - countries and states that were impacted most severely during the first wave have managed the progression from cases to deaths more successfully in subsequent waves.

\begin{table}
\begin{tabular}{ |p{2.9cm}||p{1cm}|p{1cm}|p{1cm}|p{1cm}|p{1cm}|}
 \hline
 \multicolumn{6}{|c|}{European country offsets and mortality ratios} \\
 \hline
 Country & $\tau$ & $M_1/M_2$ & $\lambda_1$ & $\lambda_2$ & $N_1/N_2$\\
 \hline
 Albania & 13 & 1.25 & 0 & 13 & 1.18 \\
 Andorra & 6 & 8.80 & 3 & 6  & 8.38 \\
 Austria & 11 & 4.08 & 11 & 11 & 4.08 \\
 Belarus & 12 & 0.72 & 18 & 5 & 0.88 \\
 Belgium & 13 & 11.25 & 4 & 16 & 10.78 \\
 Bosnia-Herzegovina & 12 & 1.92 & 3 & 12 & 1.83 \\
 Bulgaria & 13 & 1.70 & 7 & 13 & 1.57 \\
 Croatia & 8 & 2.70 & 17 & 8 & 2.87 \\
 Czech Republic & 12 & 2.19 & 10 & 12 & 2.18 \\
 Denmark & 3 & 14.28 & 0 & 27 & 7.90 \\
 Estonia & 18 & 4.57 & 10 & 10 & 6.23 \\
 Finland & 16 & 8.45 & 16 & 9 & 9.67 \\
 France & 18 & 13.67 & 5 & 18 & 13.59 \\
 Germany & 19 & 3.62 & 12 & 12 & 4.60 \\
 Greece & 16 & 2.08 & 5 & 16 & 1.99 \\
 Hungary & 4 & 6.25 & 6 & 4 & 6.25 \\
 Iceland & 26 & 1.02 & 11 & 26 & 1.02 \\
 Ireland & 9 & 9.25 & 9 & 18 & 9.14 \\
 Italy & 19 & 5.22 & 5 & 12 & 7.33 \\
 Latvia & 19 & 1.05 & 22 & 19 & 1.09 \\
 Lithuania & 11 & 3.66 & 12 & 11 & 3.78 \\
 Luxembourg & 11 & 3.69 & 7 & 11 & 3.66 \\
 Macedonia & 4 & 1.99 & 12 & 4 & 2.38 \\
 Malta & 18 & 0.76 & 2 & 22 & 0.47 \\
 Moldova & 12 & N/A & N/A & N/A &  N/A \\
 Montenegro & 7 & 1.44 & 9 & 7 & 1.44 \\
 Netherlands & 11 & 16.17 & 4 & 11 & 16.17 \\
 Norway & 20 & 6.26 & 17 & 9 & 8.80 \\
 Poland & 13 & 2.07 & 4 & 13 & 1.98 \\
 Portugal & 10 & 3.15 & 5 & 10 & 3.00 \\
 Romania & 26 & 1.95 & 4 & 26 & 1.61 \\
 Serbia & 11 & 1.42 & 0 & 11 & 1.37 \\
 Slovakia & 12 & 2.26 & 9 & 12 & 2.26 \\
 Slovenia & 12 & 5.69 & 10 & 12 & 5.64 \\
 Spain & 6 & 9.52 & 6 & 13 & 8.77 \\
 Sweden & 28 & 3.77 & 0 & 1 & 12.11 \\
 Switzerland & 13 & 5.12 & 11 & 13 & 5.11 \\
 Ukraine & 5 & N/A & N/A & N/A & N/A \\
 United Kingdom & 12 & 9.50 & 6 & 20 & 7.72 \\
\hline
\end{tabular}
\caption{European countries and their estimated offsets (which estimate the average time delay between cases and deaths) and mortality ratios (which measure the reduction in mortality between first wave and subsequent period), as defined in Section \ref{sec:secondsurgemethod}. A substantial heterogeneity in mortality ratios is observed. Generally, the one- and two-offset models produce similar results.}
\label{tab:Europe_offsets}
\end{table}

\begin{table}
\begin{tabular}{ |p{2.25cm}||p{1cm}|p{1cm}|p{1cm}|p{1cm}|p{1cm}|}
 \hline
 \multicolumn{6}{|c|}{U.S. state offsets and mortality ratios} \\
 \hline
 State & $\tau$ & $M_1/M_2$ & $\lambda_1$ & $\lambda_2$ & $N_1/N_2$ \\
 \hline
 Alabama & 18 & 1.13 & 4 & 18  & 1.06 \\
 Alaska & 29 & 3.36 & 3 & 16  & 4.07 \\
 Arizona & 13 & 1.39 & 13 & 20  & 1.26 \\
 Arkansas & 11 & 0.69 & 25 & 26  & 0.72 \\
 California & 13 & 1.62 & 6 & 27  & 1.36 \\
 Colorado & 25 & 3.81 & 5 & 5  & 7.01 \\
 Connecticut & 13 & 6.67 & 6 & 13  & 6.62 \\
 Delaware & 6 & 2.52 & 9 & 21  & 2.66 \\
 D.C. & 6 & 4.27 & 6 & 19  & 4.10 \\
 Florida & 19 & 2.84 & 11 & 19  & 2.61 \\
 Georgia & 29 & 1.30 & 13 & 3  & 1.30 \\
 Hawaii & 27 & 1.90 & 8 & 27  & 1.90 \\
 Idaho & 13 & 3.21 & 11 & 13  & 3.12 \\
 Illinois & 19 & 3.42 & 5 & 12  & 4.00 \\
 Indiana & 18 & 3.76 & 4 & 11  & 4.29 \\
 Iowa & 18 & 2.52 & 7 & 18  & 2.40 \\
 Kansas & 6 & 2.24 & 3 & 6  & 2.20 \\
 Kentucky & 13 & 3.91 & 5 & 27  & 3.06 \\
 Louisiana & 13 & 3.63 & 11 & 20  & 3.56 \\
 Maine & 6 & 2.79 & 9 & 10  & 2.51 \\
 Maryland & 11 & 3.81 & 4 & 4  & 4.08 \\
 Massachusetts & 12 & 2.49 & 4 & 6  & 2.80 \\
 Michigan & 20 & 4.78 & 6 & 13  & 5.85 \\
 Minnesota & 12 & 4.15 & 4 & 12  & 3.99 \\
 Mississippi & 13 & 1.62 & 6 & 13  & 1.56 \\
 Missouri & 40 & N/A & N/A & N/A  & N/A \\
 Montana & 14 & 2.40 & 3 & 14  & 2.10 \\
 Nebraska & 13 & 1.63 & 5 & 13  & 1.55 \\
 Nevada & 18 & 1.65 & 18 & 13  & 1.82 \\
 New Hampshire & 26 & 3.13 & 12 & 26  & 3.09 \\
 New Jersey & 11 & 8.23 & 11 & 13  & 7.51 \\
 New Mexico & 13 & 1.88 & 5 & 13  & 1.81 \\
 New York & 5 & 7.41 & 2 & 17  & 5.97 \\
 North Carolina & 4 & N/A & N/A & N/A  & N/A \\
 North Dakota & 47 & 1.00 & 11 & 27  & 1.32 \\
 Ohio & 26 & 3.93 & 9 & 19  & 3.91 \\
 Oklahoma & 20 & 5.86 & 6 & 20  & 5.31 \\
 Oregon & 11 & 2.97 & 4 & 11  & 2.89 \\
 Pennsylvania & 20 & 3.40 & 11 & 7  & 4.41 \\
 Rhode Island & 27 & 2.83 & 6 & 27  & 2.71 \\
 South Carolina & 19 & 2.95 & 18 & 19  & 2.89 \\
 South Dakota & 13 & N/A & N/A & N/A  & N/A \\
 Tennessee & 3 & 0.88 & 19 & 3  & 1.03 \\
 Texas & 20 & 1.36 & 20 & 7  & 1.57 \\
 Utah & 12 & 1.97 & 4 & 12  & 1.88 \\
 Vermont & 19 & 9.17 & 11 & 6  & 12.54 \\
 Virginia & 11 & 2.06 & 1 & 18  & 1.81 \\
 Washington & 13 & 4.09 & 3 & 13  & 3.49 \\
 West Virginia & 12 & 2.00 & 12 & 12  & 2.00 \\
 Wisconsin & 6 & 1.94 & 6 & 6  & 1.94 \\
 Wyoming & 12 & 2.09 & 11 & 12  & 2.09 \\
\hline
\end{tabular}
\caption{U.S. states and their estimated offsets and mortality ratios, as defined in Section \ref{sec:secondsurgemethod}. Missouri, North Carolina and South Dakota are determined to be in their first wave, and so do not have values of $T_1, M_i,N_i$ or $\lambda_i$. Northeastern states generally exhibit higher mortality ratios. The one- and two-offset models produce similar results.}
\label{tab:USA_offsets}
\end{table}

To understand this phenomenon further, we divided the European countries into two similarly sized groups based on the total number of cases in the first wave, adjusted for the population. As with Table \ref{tab:Europe_offsets}, we exclude the three smallest European countries Liechtenstein, Monaco and San Marino, in addition to Moldova and Ukraine, which are determined to still be in their first wave. The group that experienced a more substantial first wave (more than 1.5 total first wave cases per 1000 of population) consisted of 18 countries: Andorra, Austria, Belarus, Belgium, Denmark, France, Germany, Italy, Iceland, Ireland, Luxembourg, the Netherlands, Norway, Portugal, Spain, Sweden, Switzerland, and the U.K. The second group, which experienced a smaller first wave  (less than 1.5 cases per 1000 population), was comprised of the following 19 countries: Albania, Bosnia and Herzegovina, Bulgaria, Croatia, Czech Republic, Estonia, Finland, Greece, Hungary, Latvia, Lithuania, Macedonia, Malta, Montenegro, Poland, Serbia, Slovenia, and Slovakia. The average $N_1/N_2$ ratio for the first group was 7.39, while the same ratio for the second group was 2.94, illustrating that the first group had a more substantial decrease in the mortality rate between the first wave and subsequent period. We confirmed the statistical significance of the aforementioned difference in the $N_1/N_2$ ratio via a two-sample t-test, which had a $p$-value of 0.0005. We conducted a similar analysis for the U.S. states; however, we did not reach the same conclusions.  In fact, the second group of U.S. states (which experienced a smaller first wave of cases) had a slightly greater decrease in the mortality rate than the first group of states (the difference was not statistically significant). We observed the same relationship for a wide range of threshold values (between 1.5 and 20) for determining the partition into two groups of states.

Finally, we include a brief statistical analysis of the offsets. More homogeneity among the countries is observed in the two offsets $\lambda_1$ and $\lambda_2$, with $\lambda_2$ systematically larger than $\lambda_1$ among both European countries and U.S. states. The average offset difference, that is, $\lambda_2-\lambda_1$, across the European countries is 4.89.  The statistical significance of this observed difference is supported by the paired $t$-test, which yielded a $p$-value of 0.00183. Similarly, the corresponding average offset difference across the U.S. states is 6.48.  The statistical significance of the difference is again confirmed by the paired $t$-test, whose $p$-value was less than 0.00001.

\section{Trajectory analysis}
\label{sec:Trajectories}

This section seeks to analyze and classify countries according to appropriately normalized trajectories of their case and death counts. Let $x_i(t), y_i(t)$ be the multivariate time series of cases and deaths across a collection of countries and states. We consider all the European countries and U.S. states in conjunction, giving us a collection of size $n=93$. We normalize these time series in three ways.

Let $||\textbf{x}_i||=\sum_{t=0}^T x_i(t)$ be the $L^1$ norm of $x_i(t)$, understood as a vector in $\mathbb{R}^{T+1}$, and analogously $||\textbf{y}_i||$. Let $\textbf{c}_i=\frac{\textbf{x}_i}{||\textbf{x}_i||}$. This vector reflects the changes of the daily case time series for a given country across the entire period of analysis. Let $\textbf{d}_i=\frac{\textbf{y}_i}{||\textbf{y}_i||}$ be the normalized death time series. We define the \emph{trajectory distance matrix} $D_{ij}=||\mathbf{c}_i-\mathbf{c}_j||+||\mathbf{d}_i-\mathbf{d}_j||$ that measure distance between  normalized trajectories. Note that all vectors $\mathbf{c}_i,\mathbf{c}_j,\mathbf{d}_i,\mathbf{d}_j$ have norm 1. So a comparison is appropriate.

We may also normalize the death time series in a different way, relative to total cases. Let $\mathbf{r}_i=\frac{\textbf{y}_i}{||\textbf{x}_i||}$. This normalizes a country's trajectory of deaths according to the total number of observed cases; it captures differences not in just the trajectory of cases but separates countries more according to their overall mortality rate. We define the \emph{trajectory rate matrix} by $R_{ij}=||\mathbf{r}_i-\mathbf{r}_j||.$

\begin{figure*}
    \centering
    \begin{subfigure}[b]{0.99\textwidth}
        \includegraphics[width=\textwidth]{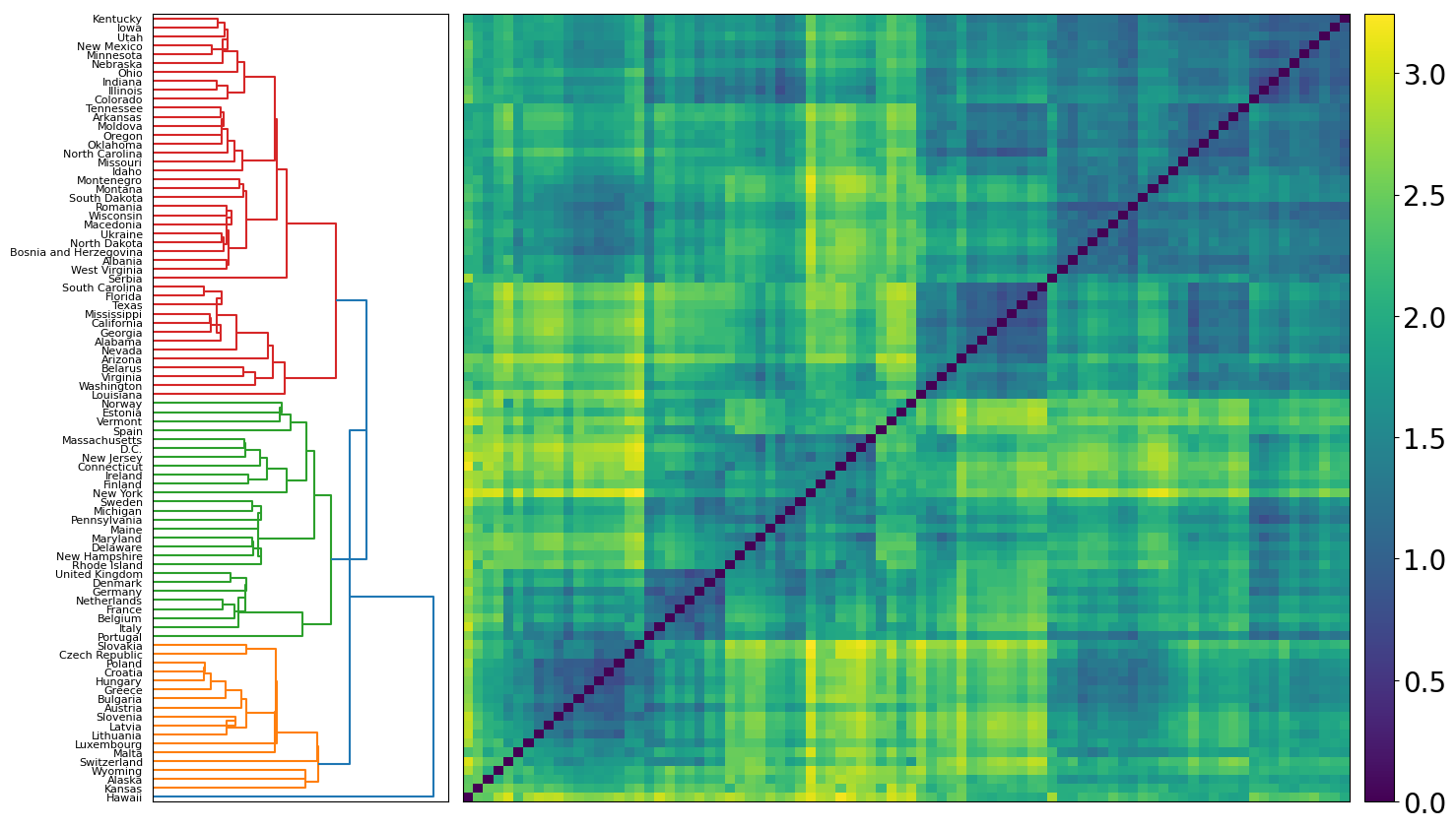}
    \end{subfigure}
    \caption{Hierarchical clustering on the matrix $D$, defined in Section \ref{sec:Trajectories}. This groups countries and states according to their overall similarity in normalized COVID-19 trajectories. The five smallest European countries, which initially appeared as outliers, have been removed. The remaining dendrogram exhibits three characteristic clusters. One consists of predominantly wealthy Western European countries and Northeastern U.S. states. One consists predominantly of less affluent European countries. The largest consists predominantly of the remaining U.S. states. The similarity between Western European countries and Northeastern U.S. states is greater than that with their respective neighboring countries and states.}
    \label{fig:Distance_dendrogram}
\end{figure*}

We can now analyze the hierarchical clustering of the $93 \times 93$ matrices $D$ and $R$. Clustering based on $R$ reveals New York, New Jersey, Connecticut and Massachusetts as clear outliers. Indeed, these four U.S. states featured very high mortality rates in the early days of the pandemic in the United States. Clustering on $D$ reveals far more insights. In the first instance, a cluster of Monaco, Liechtenstein, Iceland, Andorra and San Marino arises as clear outliers. We have removed these countries, all the five smallest in Europe, to obtain Figure \ref{fig:Distance_dendrogram}.

Turning now to a close analysis of Figure \ref{fig:Distance_dendrogram}, we observe the existence of three distinct clusters. The green cluster is almost exclusively composed of Northeastern U.S. states (Connecticut, Delaware, D.C., Maine, Maryland, Massachusetts, New Hampshire, New Jersey, New York, Pennsylvania, Rhode Island, and Vermont) and developed Western or Northern European countries (Belgium, Denmark, Finland, France, Germany, Ireland, Italy, the Netherlands, Norway, Spain, Sweden, and the U.K.). As explored in Section \ref{sec:offset}, many of these experienced similar severe first waves, with most responding with lockdowns. These countries and states managed their progression from cases to deaths much more successfully in subsequent waves of COVID-19.

The orange cluster consists primarily of less developed European countries (Bulgaria, Croatia, Czech Republic, Greece, Hungary, Latvia, Lithuania, Poland, Slovakia and Slovenia) and select U.S. states such as Alaska and Kansas. These countries and states all experienced less severe first waves, and more significant subsequent waves of the virus. Countries such as Croatia and Greece continued to attract travelers during the European summer, \cite{Greecetourism} which may have been an additional factor in spreading the virus. U.S. states such as Alaska and Kansas also both experienced less severe first waves in COVID-19, followed by extreme growth in both cases and deaths in subsequent waves. 

The red cluster is composed mostly of the remaining U.S. states and select European countries. Their case and death count have been steadily increasing over the entire period, for the most part.

\section{Conclusion}
\label{sec:Conclusion}
This paper introduces new methods to partition time series and estimate offsets to quantify the changing mortality of COVID-19 over different waves of the pandemic. The methodology is applied to European countries and U.S. states, both independently and in conjunction. Our methodology is flexible: different smoothing techniques, metrics between data, parameters in the algorithmic framework, and clustering methods can be used to study multivariate time series and identify changing mortality or other outcomes beyond this application.

Our analysis has refined the popular conception concerning the reduction of mortality during the second wave of COVID-19 in Europe.\cite{FTsecondsurge} We have shown significant variance in the mortality ratios (defined in Section \ref{sec:secondsurgemethod}), with European countries spanning a wide spectrum. For example, the Netherlands has reduced its mortality drastically, Germany has done so moderately, while  Belarus' mortality has slightly increased. Wealthy Western and Northern European countries, with the notable exceptions of Germany and Sweden, have reduced their mortality more than all U.S. states and the rest of Europe. Similar findings are observed with both one- and two-offset models. All these findings are reflected in the scatter plots of Figure \ref{fig:Convex_hull_plots}, where a great difference in time-adjusted case and death data points is observed between the first wave and subsequent period. We remark that this subsequent period may consist of either a single second wave, or a second and third wave, and in the latter case, we consider the second and third waves together in one unit we call the subsequent period. Our motivation for this is the hypothesis that the first wave was of an exceptionally different nature than subsequent waves, being a time in which a new disease took Europe by surprise, and there were fewer effective treatments known.

This considerable reduction in mortality, as observed with our methodology, has several explanations. In Europe, it has been surmised that the first wave's case counts were drastically underestimated. \cite{underreporting} The first wave disproportionately affected the elderly, \cite{Kontopantelis2020} while the second wave has largely affected young people, \cite{Aleta2020} who have a much lower morbidity rate. It is also possible that the more socialized and equitable health systems of wealthy European countries have served their populations better than the United States, \cite{UShealthcare} resulting in higher mortality reductions than even the Northeastern U.S. states.

These various interpretations of mortality reduction also reveal some limitations of this article. Not only were many early cases under-reported, \cite{underreporting} but testing protocols have been far from uniform over time and between countries. Indeed, several countries have changed their testing protocols on various occasions, including within the same wave. \cite{Francechange,Pullano2020,Antigenchange} Within Italy, for example, different regions operated according to different protocols, testing only symptomatic patients or more broadly. \cite{DiBari2020} Even deaths may have been under-reported, with substantial differences having been observed between excess mortality and reported COVID-19 deaths. \cite{italydeaths} To complement our analysis, we also applied our methodology to two different time series: estimated true COVID-19 cases (as calculated by the IHME model \cite{IHMEmodel}) and excess mortality data. We implement this and discuss its findings and limitations in Appendix \ref{appendix:excessmortality}.

Despite the considerable variance in mortality reduction among European countries and the U.S. states, we have also shown broad similarity between the two groups. Hierarchical clustering on the one- and two- offset mortality pairs, shown in Figure \ref{fig:Dendrogram_M1M2}, highlights a broadly similar structure among Europe and the United States. Both groups, for the one- and two- offset models, consist of one predominant cluster and a smaller cluster. In each group, the smaller cluster consists of countries and states that experienced a severe first wave and were then able to substantially reduce their mortality during the subsequent period. These were predominantly wealthy Western European countries and Northeastern U.S. states. A consistent observation among both groups is that countries and states that experienced a severe first wave in cases reduced their mortality rate more effectively in subsequent waves. We then confirmed this in a statistical test for Europe, although it failed for the United States.

Our offset analysis also revealed insights regarding the time delay between cases and deaths in the first and subsequent waves. This finding was more consistent than the mortality reduction across both Europe and the United States. We found that the time offset $\lambda_2$ between cases and deaths of the subsequent period is systematically larger than that of the first wave, $\lambda_1$. This is likely due to under-reporting in the first wave. Values of $\lambda_1$ as small as zero indicate that during the first waves, the spike in deaths was happening at the exact same time as the spike in cases, suggesting that the cases were being observed too late, or not at all. Indeed, previous analysis \cite{chaos1James2020} has showed that Spain had 2.27 times as many deaths on 03/28/2020 as the number of cases 16 days earlier. By the second wave, European countries and U.S. states were testing more consistently, so $\lambda_2$ is likely a more accurate reflection of the time delay between cases and deaths. The same limitations as previously stated apply, due to countries changing their testing protocols even within the second wave, and varying between countries. \cite{Antigenchange}

Finally, Figure \ref{fig:Distance_dendrogram} analyzed all the European countries and U.S. states in conjunction, and again highlighted numerous similarities and differences between the groups. One cluster consisted of most wealthy Western European countries and Northeastern states alike, indicating their similar trajectories of both cases and deaths. Another cluster consists primarily of less affluent European countries such as Bulgaria, Croatia, Greece, Latvia and Poland. The third cluster consists primarily of the remaining U.S. states. In particular, while similarity exists between Western European countries and Northeastern U.S. states, there is no such close relationship between other European countries, such as less developed Eastern European countries, and other U.S. states outside the Northeast.

Overall, this paper introduces a new method for analyzing second wave mortality in a collection of epidemiological time series and provides new insights into the differing effects of COVID-19 across Europe and the U.S. Midway through 2020, Europe had seen a substantial reduction of new case counts at the end of its first wave, and many European countries were praised for their handling of the virus. Few predicted the enormity of the second wave in Europe. Fortunately, wealthy Western countries, the Netherlands most of all, experienced a drastic drop in mortality. However, this is by no means uniform. Less developed countries are often underrepresented in media reports, and their mortality is less visible in popular conception. As further waves carry the risk of widespread loss of life, each country must be aware of the potentially high human cost of COVID-19 and react swiftly to new waves of the pandemic. Through research, cooperation and learning from each other's successes, European countries may reduce the human and other costs of the pandemic.\cite{Momtazmanesh2020,Priesemann2021}

\section*{Data availability}
The data that support the findings of this study are openly available at Refs. \onlinecite{worldindata2020} and \onlinecite{datasource}.
\begin{acknowledgments}
The authors thank Kerry Chen and Jarrah Lacko for helpful comments and edits.
\end{acknowledgments}

\appendix

\section{Turning point methodology}
\label{appendix:turningpoint}

In this section, we provide more details for the identification of turning points of a new case time series $x(t)$, in particular the first non-trivial trough $T_1$. First, some smoothing of the time series is necessary due to irregularities in the data set, and discrepancies between different data sources. There are lower counts on the weekends, and some negative counts due to retroactive adjustments. A Savitzy-Golay filter ameliorates these issues by combining polynomial smoothing with a moving average computation - this moving average eliminates all but a few small negative counts; we then replace these negative counts with zero. This yields a smoothed time series $\hat{x}(t) \in \mathbb{R}_{\geq 0}.$ Subsequently, we perform a two-step process to select and then refine a non-empty set $P$ of local maxima (peaks) and $T$ of local minima (troughs). Then $T_1$ is the first non-trivial element of $T$.

Following Ref. \onlinecite{james2020covidusa}, we apply a two-step algorithm to the smoothed time series $\hat{x}(t)$. The first step produces an alternating sequence of troughs and peaks, beginning with a trough at $t=0$, where there are zero cases. The second step refines this sequence according to chosen conditions and parameters. The most important conditions to initially identify a peak or trough, respectively, are the following:
\begin{align}
\label{baddefnpeak}
\hat{x}(t_0)&=\max\{\hat{x}(t): \max(0,t_0 - l) \leq t \leq \min(t_0 + l,T)\},\\
\label{baddefntrough}\hat{x}(t_0)&=\min\{\hat{x}(t): \max(0,t_0 - l) \leq t \leq \min(t_0 + l,T)\},
\end{align}
where $l$ is a parameter to be chosen. Following Ref. \onlinecite{james2020covidusa}, we select $l=17$, which accounts for the 14-day incubation period of the virus \cite{incubation2020} and less testing on weekends. Defining peaks and troughs according to this definition alone has some flaws, such as the potential for two consecutive peaks.

Instead, we implement an inductive procedure to choose an alternating sequence of peaks and troughs. Suppose $t_0$ is the last determined peak. We search in the period $t>t_0$ for the first of two cases: if we find a time $t_1>t_0$ that satisfies (\ref{baddefntrough}) as well as a non-triviality condition $\hat{x}(t_1)<\hat{x}(t_0)$, we add $t_1$ to the set of troughs and proceed from there. If we find a time $t_1>t_0$ that satisfies (\ref{baddefnpeak}) and  $\hat{x}(t_0)\geq \hat{x}(t_1)$, we ignore this lower peak as redundant; if we find a time $t_1>t_0$ that satisfies (\ref{baddefnpeak}) and  $\hat{x}(t_1) > \hat{x}(t_0)$, we remove the peak $t_0$,  replace it with $t_1$ and continue from $t_1$. A similar process applies from a trough at $t_0$.  %

At this point, the time series is assigned an alternating sequence of troughs and peaks. However, some turning points are immaterial and should be removed. Let $t_1<t_3$ be two peaks, necessarily separated by a trough. We select a parameter $\delta=0.2$, and if the \emph{peak ratio}, defined as $\frac{\hat{x}(t_3)}{\hat{x}(t_1)}<\delta$, we remove the peak $t_3$. If two consecutive troughs $t_2,t_4$ remain, we remove $t_2$ if $\hat{x}(t_2)>\hat{x}(t_4)$, otherwise remove $t_4$. That is, if the second peak has size less than $\delta$ of the first peak, we remove it. 

Finally, we use the same \emph{log-gradient} function between times $t_1<t_2$, defined as
\begin{align}
\label{loggrad}
   \loggrad(t_1,t_2)=\frac{\log \hat{x}(t_2) - \log \hat{x}(t_1)}{t_2-t_1}.
\end{align}
The numerator equals  $\log(\frac{\hat{x}(t_2)}{\hat{x}(t_1)})$, a "logarithmic rate of change." Unlike the standard rate of change given by $\frac{\hat{x}(t_2)}{\hat{x}(t_1)} -1$, the logarithmic change is symmetrically between $(-\infty,\infty)$. Let $t_1,t_2$ be adjacent turning points (one a trough, one a peak). We choose a parameter $\epsilon=0.01$;  if
\begin{align}
    |\loggrad(t_1,t_2)|<\epsilon,
\end{align}
that is, the average logarithmic change is less than 1\%, we remove $t_2$ from our sets of peaks and troughs. If $t_2$ is not the final turning point, we also remove $t_1$. 

We conclude with an alternating sequence of peaks and troughs, beginning with a trough at $t=0$. We then simply define $T_1$ to be the first trough after $t=0$, if this exists. This marks the end of the first wave. We include an algorithmic presentation of our two-step procedure to determine peaks and troughs (and isolate $T_1$ in particular) in Algorithms \ref{alg:stage1} and \ref{alg:stage2}.

 \begin{algorithm}[H]
 	\begin{algorithmic} 
 	 	\caption{Turning point identification (step 1)}
 	 \label{alg:stage1}
 	\State Given: a time series $x(t) \in \mathbb{R}$
	\State Form a smoothed time series: $\hat{x}(t)$ = Savitzky-Golay$(x(t))$;
 	\State Data preprocessing: \textbf{If}  {$\hat{x}(t) < 0,  \textbf{ then } \hat{x}(t) = 0$};
	\State Initialize: state = TroughState, Current TP = 1, PeakSet = empty, TroughSet = \{1\};
    \While {Current TP $<T$}
    \State Set $t_0$ = Current TP; Flag = false;
    \For {$t_1=t_0 + 1$ to $T$}
    \If {state = TroughState \textbf{and} $t_1$ satisfies (\ref{baddefnpeak}) \textbf{and} $\hat{x}(t_1)>\hat{x}(t_0)$}
    \State state = PeakState;
    \State Append $t_1$ to PeakSet;
    \State Current TP = $t_1$; Flag = true;
    \State \textbf{Break for}
    
    \ElsIf{state = TroughState \textbf{and} $t_1$ satisfies (\ref{baddefntrough}) \textbf{and} $\hat{x}(t_1)<\hat{x}(t_0)$}
    \State Append $t_1$ to TroughSet;
    \State Remove $t_0$ from TroughSet;
    \State Current TP = $t_1$; Flag = true;
    \State \textbf{Break for}
    \ElsIf{state = PeakState \textbf{and} $t_1$ satisfies (\ref{baddefntrough}) \textbf{and} $\hat{x}(t_1)<\hat{x}(t_0)$}
    \State state = TroughState;
    \State Append $t_1$ to TroughSet;
    \State Current TP = $t_1$; Flag = true;
    \State \textbf{Break for}
     \ElsIf{ state = PeakState \textbf{and} $t_1$ satisfies (\ref{baddefnpeak}) \textbf{and} $\hat{x}(t_1)>\hat{x}(t_0)$}
    \State Append $t_1$ to PeakSet;
    \State Remove $t_0$ from PeakSet;
    \State Current TP = $t_1$; Flag = true;
    \State \textbf{Break for}
    \EndIf
    \EndFor
    \If {Flag = false}
    \State \textbf{Break while}
    \EndIf
    \EndWhile
	\State Output PeakSet and TroughSet.
   \algstore{myalg}
    \end{algorithmic}
    \end{algorithm}

  \begin{algorithm}[H]
    \begin{algorithmic}   
    \caption{Turning point refinement (step 2)}
 	 \label{alg:stage2}
    \algrestore{myalg}
  	\State TPSet = Sort(PeakSet $\cup$ TroughSet); \Comment{Indexing begins from 1} 
 	\State Initialize: CurrentPeakIndex = 2; \Comment{Begin the peak ratio refinement}
 	\While{CurrentPeakIndex $\leq$ Length(TPSet) - 2}
 	\State $i$ = CurrentPeakIndex, $t_1$ = TPSet($i$), $t_3$ = TPSet($i+2$);
\If {$\frac{\hat{x}(t_3)}{\hat{x}(t_1)} \geq \delta$}
        \State CurrentPeakIndex = $i+2$;
 \ElsIf {$\frac{\hat{x}(t_3)}{\hat{x}(t_1)} < \delta$ \textbf{and} $i+2$ = Length(TPSet)}
 \State Remove $t_3$ from PeakSet;
\State        TPSet=Sort(PeakSet $\cup$ TroughSet);
\ElsIf{ $\frac{\hat{x}(t_3)}{\hat{x}(t_1)} < \delta$ \textbf{and} $i+2 <$ Length(TPSet)}
\State $t_2$ = TPSet($i+1$), $t_4$ = TPSet($i+3$);
    \If {$\hat{x}(t_2) \leq \hat{x}(t_4)$} 
\State Remove $t_4$ from TroughSet;
\Else 
\State Remove $t_2$ from TroughSet;
\EndIf
\State Remove $t_3$ from PeakSet;
\State        TPSet = Sort(PeakSet $\cup$ TroughSet);
\EndIf
 	\EndWhile

\State Initialize: CurrentIndex = 1; \Comment{Begin the log-grad refinement}
\While{CurrentIndex $<$ Length(TPSet)}
\State $i$ = CurrentIndex, $t_0$ = TPSet($i$), $t_1$ = TPSet($i+1$);
\If{$|\loggrad(t_0, t_1)|<\epsilon$} \Comment{See Equation (\ref{loggrad})}
\State Remove $t_0$ and $t_1$ from both TroughSet and PeakSet;
\State TPSet = Sort(PeakSet $\cup$ TroughSet);
\Else 
\State CurrentIndex = $i+1$;
\EndIf
\EndWhile
\State Output PeakSet and TroughSet.
\State $T_1$ = PeakSet(2).
 	
    \end{algorithmic}
    \end{algorithm}
    
\section{Alternative data analysis}
\label{appendix:excessmortality}

In this brief section, we apply our methodology to an alternative data set. We substitute new daily reported deaths with weekly excess mortality, which is only available on a weekly basis. This is defined as the difference between reported deaths from all causes in a given week of 2020 and the average of the deaths in the same week across the five preceding years. We substitute new daily reported cases with estimated weekly cases, as calculated by the IHME model.\cite{IHMEmodel} Our data spans (the week ending) 01/26/2020 to (the week ending) 11/29/2020.

The first issue with the excess mortality data set is that the data are frequently negative. Applying our methodology in Section \ref{sec:secondsurgemethod}, where deaths are summed over the determined first wave, we sometimes produce a negative value of $M_1$. As such, we apply our methodology in two slightly different ways. First, we make no alterations to the excess mortality time series to obtain an offset $\tau$ (measured in weeks) and a one-offset mortality ratio $M_1/M_2$. Next, we replace all negative counts in the excess mortality time series with zero to obtain an offset we call $\tau'$ (in weeks) and a modified one-offset mortality ratio  $M'_1/M'_2$. We record these results for 15 European countries in Table \ref{tab:newexperiment}.

\begin{table}
\begin{tabular}{ |p{2.9cm}||p{1.1cm}|p{1.1cm}|p{1.1cm}|p{1.1cm}|}
 \hline
 \multicolumn{5}{|c|}{European country offsets and mortality ratios} \\
 \hline
 Country & $\tau$ & $M_1/M_2$ & $\tau'$ &  $M'_1/M'_2$\\
 \hline
 Austria &  1 & 1.82 & 1 & 2.36   \\
 Belgium & 1 & 1.22 & 1 & 1.53   \\
 Denmark & 1 & -0.13 & 1 & 0.60  \\
 Estonia & 2 & 0.42 & 2 & 0.91   \\
 Finland & 1 & 0.40 & 1 & 0.43   \\
 France & 1 & 1.13 & 1 & 1.35  \\
 Germany & 1 & 0.27 & 1 & 1.00  \\
 Hungary & 2 & -2.25 & 2 & 0.98  \\
 Lithuania & 3 & -1.02 & 2 & 1.89   \\
 Netherlands & 1 & 1.11 & 1 & 1.26   \\
 Norway & 1 & -0.98 & 1 & 0.87   \\
 Portugal & 1 & 1.84 & 1 & 2.27  \\
 Spain & 1 & 1.14 & 1 & 1.26  \\
 Sweden & 2 & 7.17 & 2 & 2.80   \\
 United Kingdom & 2 & 1.98 & 2 & 1.95 \\
\hline
\end{tabular}
\caption{Estimated offsets and mortality ratios for select European countries. These are calculated from weekly estimated true COVID-19 cases and excess mortality data. Mortality ratios $M_1/M_2$ may be negative due to negative counts in excess mortality, so we repeat the analysis by nullifying all negative counts to produce $M'_1/M'_2$. Offsets are calculated and measured in weeks.}
\label{tab:newexperiment}
\end{table}

There are several findings captured here and limitations to consider. First, using excess mortality data produces a negative value of $M_1$, and hence a negative mortality ratio, for four countries. Indeed, the decrease in mortality in 2020 relative to the last five years was greater than the number of COVID-19 deaths during these countries' first waves. Substantial negative excess mortality counts in January and February contributed to this.

As no week should contribute a negative number of deaths, we repeated the analysis by nullifying any negative counts in excess mortality, to obtain ratios $M'_1/M'_2$. This also has a drawback - during the middle of the year, there are still several weeks with negative excess mortality counts, and so this nullification systematically raises the number of deaths in the observed excess mortality counts. We executed the analysis a third time by selectively nullifying only negative excess mortality counts in January and February, during which there were relatively few COVID-19 deaths, and obtained similar results as $M'_1/M'_2$.

In general, excess mortality is a problematic estimate of COVID-19 mortality because it considers all deaths from all causes. Responding to the pandemic, people's regular behavior has substantially changed in 2020. Flu infections substantially decreased, \cite{fludeaths} as did traffic, \cite{Qureshi2020,Saladi2020} although its effect on mortality has been unclear. The widespread impacts of the pandemic on mental health have been researched, with some indication that suicides may have increased or decreased relative to previous years, depending on the country. \cite{John2020,Le2020}

With all this in mind, Table \ref{tab:newexperiment} contains several insights. First, the offsets $\tau$ and $\tau'$ agree for all but one country, and overwhelmingly suggest an offset of one to two weeks, mirroring the results of Table \ref{tab:Europe_offsets}. In addition, most modified mortality ratios $M'_1/M'_2$ are greater than 1, highlighting a decrease in mortality between the first wave and subsequent period. Of the five countries with $M'_1/M'_2 < 1$, three have $M_1<0$, suggesting that nullifying negative excess mortality counts has added a large number of deaths to the total. Thus, these ratios may not be reliable. The differences between the mortality ratios in Tables \ref{tab:Europe_offsets} and \ref{tab:newexperiment} suggests that the underestimation of true cases in the first wave plays a substantial role in the large reduction of mortality with respect to reported cases and deaths, as discussed in the body of the paper.

Overall, the analysis performed in Section \ref{sec:offset} is not without its limitations. The substantial decrease in mortality according to reported cases and deaths has several explanations, including potential better treatments or underestimation of true COVID-19 cases, or even deaths - this appendix suggests the latter plays a more substantial role. Further research is needed, including incorporating hospitalization data, to reveal the true changes in mortality from COVID-19 in different ways of the disease. Future work could also study the difference between the second and third waves more closely - we have only distinguished the first wave versus subsequent period. This research may assist in an effective, coordinated European response to the pandemic.

\bibliography{_references}

\begin{thebibliography}{63}%
\makeatletter
\providecommand \@ifxundefined [1]{%
 \@ifx{#1\undefined}
}%
\providecommand \@ifnum [1]{%
 \ifnum #1\expandafter \@firstoftwo
 \else \expandafter \@secondoftwo
 \fi
}%
\providecommand \@ifx [1]{%
 \ifx #1\expandafter \@firstoftwo
 \else \expandafter \@secondoftwo
 \fi
}%
\providecommand \natexlab [1]{#1}%
\providecommand \enquote  [1]{``#1''}%
\providecommand \bibnamefont  [1]{#1}%
\providecommand \bibfnamefont [1]{#1}%
\providecommand \citenamefont [1]{#1}%
\providecommand \href@noop [0]{\@secondoftwo}%
\providecommand \href [0]{\begingroup \@sanitize@url \@href}%
\providecommand \@href[1]{\@@startlink{#1}\@@href}%
\providecommand \@@href[1]{\endgroup#1\@@endlink}%
\providecommand \@sanitize@url [0]{\catcode `\\12\catcode `\$12\catcode
  `\&12\catcode `\#12\catcode `\^12\catcode `\_12\catcode `\%12\relax}%
\providecommand \@@startlink[1]{}%
\providecommand \@@endlink[0]{}%
\providecommand \url  [0]{\begingroup\@sanitize@url \@url }%
\providecommand \@url [1]{\endgroup\@href {#1}{\urlprefix }}%
\providecommand \urlprefix  [0]{URL }%
\providecommand \Eprint [0]{\href }%
\providecommand \doibase [0]{https://doi.org/}%
\providecommand \selectlanguage [0]{\@gobble}%
\providecommand \bibinfo  [0]{\@secondoftwo}%
\providecommand \bibfield  [0]{\@secondoftwo}%
\providecommand \translation [1]{[#1]}%
\providecommand \BibitemOpen [0]{}%
\providecommand \bibitemStop [0]{}%
\providecommand \bibitemNoStop [0]{.\EOS\space}%
\providecommand \EOS [0]{\spacefactor3000\relax}%
\providecommand \BibitemShut  [1]{\csname bibitem#1\endcsname}%
\let\auto@bib@innerbib\@empty
\bibitem [{\citenamefont {Looi}(2020)}]{Looi2020}%
  \BibitemOpen
  \bibfield  {author} {\bibinfo {author} {\bibfnamefont {M.-K.}\ \bibnamefont
  {Looi}},\ }\bibfield  {title} {\enquote {\bibinfo {title} {Covid-19: Is a
  second wave hitting {E}urope?}}\ }\href {https://doi.org/10.1136/bmj.m4113}
  {\bibfield  {journal} {\bibinfo  {journal} {{BMJ}}\ ,\ \bibinfo {pages}
  {m4113}} (\bibinfo {year} {2020})}\BibitemShut {NoStop}%
\bibitem [{\citenamefont {Walsh}()}]{bbcnormal}%
  \BibitemOpen
  \bibfield  {author} {\bibinfo {author} {\bibfnamefont {F.}~\bibnamefont
  {Walsh}},\ }\href@noop {} {\enquote {\bibinfo {title} {Coronavirus: Is it
  time to move on and get back to normal life?}}\ }\bibinfo {howpublished}
  {\url{https://www.bbc.com/news/health-53951764 }},\ \bibinfo {note} {{BBC},
  August 28, 2020}\BibitemShut {NoStop}%
\bibitem [{\citenamefont {Cookson}\ and\ \citenamefont
  {Burn-Murdoch}()}]{FTsecondsurge}%
  \BibitemOpen
  \bibfield  {author} {\bibinfo {author} {\bibfnamefont {C.}~\bibnamefont
  {Cookson}}\ and\ \bibinfo {author} {\bibfnamefont {J.}~\bibnamefont
  {Burn-Murdoch}},\ }\href@noop {} {\enquote {\bibinfo {title} {Why the second
  wave of {C}ovid-19 appears to be less lethal},}\ }\bibinfo {howpublished}
  {\url{https://www.ft.com/content/b3801b63-fbdb-433b-9a46-217405b1109f}},\
  \bibinfo {note} {{F}inancial Times, October 21, 2020}\BibitemShut {NoStop}%
\bibitem [{\citenamefont {McDonell}()}]{bbccloseborders_2020}%
  \BibitemOpen
  \bibfield  {author} {\bibinfo {author} {\bibfnamefont {S.}~\bibnamefont
  {McDonell}},\ }\href@noop {} {\enquote {\bibinfo {title} {Coronavirus: {US}
  and {A}ustralia close borders to {C}hinese arrivals},}\ }\bibinfo
  {howpublished} {\url{https://www.bbc.com/news/world-51338899}},\ \bibinfo
  {note} {{BBC}, February 2, 2020}\BibitemShut {NoStop}%
\bibitem [{\citenamefont {McCurry}()}]{guardian_2020}%
  \BibitemOpen
  \bibfield  {author} {\bibinfo {author} {\bibfnamefont {J.}~\bibnamefont
  {McCurry}},\ }\href@noop {} {\enquote {\bibinfo {title} {Test, trace,
  contain: how {S}outh {K}orea flattened its coronavirus curve},}\ }\bibinfo
  {howpublished}
  {\url{https://www.theguardian.com/world/2020/apr/23/test-trace-contain-how-south-korea-flattened-its-coronavirus-curve}},\
  \bibinfo {note} {{The Guardian}, U.S.April 23, 2020}\BibitemShut {NoStop}%
\bibitem [{\citenamefont {McCann}, \citenamefont {Popovich},\ and\
  \citenamefont {Wu}()}]{nyt2020}%
  \BibitemOpen
  \bibfield  {author} {\bibinfo {author} {\bibfnamefont {A.}~\bibnamefont
  {McCann}}, \bibinfo {author} {\bibfnamefont {N.}~\bibnamefont {Popovich}},\
  and\ \bibinfo {author} {\bibfnamefont {J.}~\bibnamefont {Wu}},\ }\href@noop
  {} {\enquote {\bibinfo {title} {Italy’s virus shutdown came too late. what
  happens now?}}\ }\bibinfo {howpublished}
  {\url{https://www.nytimes.com/interactive/2020/04/05/world/europe/italy-coronavirus-lockdown-reopen.html}},\
  \bibinfo {note} {{The New York Times}, U.S.April 5, 2020}\BibitemShut
  {NoStop}%
\bibitem [{\citenamefont {Scally}, \citenamefont {Jacobson},\ and\
  \citenamefont {Abbasi}(2020)}]{Scally2020}%
  \BibitemOpen
  \bibfield  {author} {\bibinfo {author} {\bibfnamefont {G.}~\bibnamefont
  {Scally}}, \bibinfo {author} {\bibfnamefont {B.}~\bibnamefont {Jacobson}},\
  and\ \bibinfo {author} {\bibfnamefont {K.}~\bibnamefont {Abbasi}},\
  }\bibfield  {title} {\enquote {\bibinfo {title} {The {UK}'s public health
  response to covid-19},}\ }\href {https://doi.org/10.1136/bmj.m1932}
  {\bibfield  {journal} {\bibinfo  {journal} {{BMJ}}\ ,\ \bibinfo {pages}
  {m1932}} (\bibinfo {year} {2020})}\BibitemShut {NoStop}%
\bibitem [{\citenamefont {Iati}\ \emph {et~al.}()\citenamefont {Iati} \emph
  {et~al.}}]{wapo_allreopen}%
  \BibitemOpen
  \bibfield  {author} {\bibinfo {author} {\bibfnamefont {M.}~\bibnamefont
  {Iati}} \emph {et~al.},\ }\href@noop {} {\enquote {\bibinfo {title} {All 50
  {U.S.} states have taken steps toward reopening in time for {Memorial Day}
  weekend},}\ }\bibinfo {howpublished}
  {\url{https://www.washingtonpost.com/nation/2020/05/19/coronavirus-update-us}},\
  \bibinfo {note} {{The Washington Post}, May 20, 2020}\BibitemShut {NoStop}%
\bibitem [{\citenamefont {Meyer}\ and\ \citenamefont
  {Madrigal}()}]{atlantic_secondsurge}%
  \BibitemOpen
  \bibfield  {author} {\bibinfo {author} {\bibfnamefont {R.}~\bibnamefont
  {Meyer}}\ and\ \bibinfo {author} {\bibfnamefont {A.~C.}\ \bibnamefont
  {Madrigal}},\ }\href@noop {} {\enquote {\bibinfo {title} {A devastating new
  stage of the pandemic},}\ }\bibinfo {howpublished}
  {\url{https://www.theatlantic.com/science/archive/2020/06/second-coronavirus-surge-here/613522}},\
  \bibinfo {note} {the {A}tlantic, June 25, 2020}\BibitemShut {NoStop}%
\bibitem [{\citenamefont {James}\ and\ \citenamefont
  {Menzies}(2020{\natexlab{a}})}]{james2020covidusa}%
  \BibitemOpen
  \bibfield  {author} {\bibinfo {author} {\bibfnamefont {N.}~\bibnamefont
  {James}}\ and\ \bibinfo {author} {\bibfnamefont {M.}~\bibnamefont
  {Menzies}},\ }\bibfield  {title} {\enquote {\bibinfo {title} {{COVID}-19 in
  the {United States}: Trajectories and second surge behavior},}\ }\href
  {https://doi.org/10.1063/5.0024204} {\bibfield  {journal} {\bibinfo
  {journal} {Chaos: An Interdisciplinary Journal of Nonlinear Science}\
  }\textbf {\bibinfo {volume} {30}},\ \bibinfo {pages} {091102} (\bibinfo
  {year} {2020}{\natexlab{a}})}\BibitemShut {NoStop}%
\bibitem [{\citenamefont {Griffin}(2020)}]{Griffin2020}%
  \BibitemOpen
  \bibfield  {author} {\bibinfo {author} {\bibfnamefont {S.}~\bibnamefont
  {Griffin}},\ }\bibfield  {title} {\enquote {\bibinfo {title} {Covid-19:
  Second wave death rate is doubling fortnightly but is lower and slower than
  in march},}\ }\href {https://doi.org/10.1136/bmj.m4092} {\bibfield  {journal}
  {\bibinfo  {journal} {{BMJ}}\ ,\ \bibinfo {pages} {m4092}} (\bibinfo {year}
  {2020})}\BibitemShut {NoStop}%
\bibitem [{\citenamefont {Wang}\ \emph {et~al.}(2020)\citenamefont {Wang} \emph
  {et~al.}}]{Remdesivir}%
  \BibitemOpen
  \bibfield  {author} {\bibinfo {author} {\bibfnamefont {M.}~\bibnamefont
  {Wang}} \emph {et~al.},\ }\bibfield  {title} {\enquote {\bibinfo {title}
  {Remdesivir and chloroquine effectively inhibit the recently emerged novel
  coronavirus (2019-{nCoV}) in vitro},}\ }\href
  {https://doi.org/10.1038/s41422-020-0282-0} {\bibfield  {journal} {\bibinfo
  {journal} {Cell Research}\ }\textbf {\bibinfo {volume} {30}},\ \bibinfo
  {pages} {269--271} (\bibinfo {year} {2020})}\BibitemShut {NoStop}%
\bibitem [{\citenamefont {Bloch}(2020)}]{Bloch2020}%
  \BibitemOpen
  \bibfield  {author} {\bibinfo {author} {\bibfnamefont {E.~M.}\ \bibnamefont
  {Bloch}},\ }\bibfield  {title} {\enquote {\bibinfo {title} {Convalescent
  plasma to treat {COVID}-19},}\ }\href
  {https://doi.org/10.1182/blood.2020007714} {\bibfield  {journal} {\bibinfo
  {journal} {Blood}\ }\textbf {\bibinfo {volume} {136}},\ \bibinfo {pages}
  {654--655} (\bibinfo {year} {2020})}\BibitemShut {NoStop}%
\bibitem [{\citenamefont {Xu}\ \emph {et~al.}(2020)\citenamefont {Xu} \emph
  {et~al.}}]{toczilizumab}%
  \BibitemOpen
  \bibfield  {author} {\bibinfo {author} {\bibfnamefont {X.}~\bibnamefont {Xu}}
  \emph {et~al.},\ }\bibfield  {title} {\enquote {\bibinfo {title} {Effective
  treatment of severe {COVID}-19 patients with tocilizumab},}\ }\href
  {https://doi.org/10.1073/pnas.2005615117} {\bibfield  {journal} {\bibinfo
  {journal} {Proceedings of the National Academy of Sciences}\ }\textbf
  {\bibinfo {volume} {117}},\ \bibinfo {pages} {10970--10975} (\bibinfo {year}
  {2020})}\BibitemShut {NoStop}%
\bibitem [{\citenamefont {Cao}\ \emph {et~al.}(2020)\citenamefont {Cao} \emph
  {et~al.}}]{Cao2020}%
  \BibitemOpen
  \bibfield  {author} {\bibinfo {author} {\bibfnamefont {B.}~\bibnamefont
  {Cao}} \emph {et~al.},\ }\bibfield  {title} {\enquote {\bibinfo {title} {A
  trial of {L}opinavir-{R}itonavir in adults hospitalized with severe
  {C}ovid-19},}\ }\href {https://doi.org/10.1056/nejmoa2001282} {\bibfield
  {journal} {\bibinfo  {journal} {New England Journal of Medicine}\ }\textbf
  {\bibinfo {volume} {382}},\ \bibinfo {pages} {1787--1799} (\bibinfo {year}
  {2020})}\BibitemShut {NoStop}%
\bibitem [{\citenamefont {Li}\ \emph {et~al.}(2020)\citenamefont {Li} \emph
  {et~al.}}]{underreporting}%
  \BibitemOpen
  \bibfield  {author} {\bibinfo {author} {\bibfnamefont {R.}~\bibnamefont {Li}}
  \emph {et~al.},\ }\bibfield  {title} {\enquote {\bibinfo {title} {Substantial
  undocumented infection facilitates the rapid dissemination of novel
  coronavirus ({SARS}-{CoV}-2)},}\ }\href
  {https://doi.org/10.1126/science.abb3221} {\bibfield  {journal} {\bibinfo
  {journal} {Science}\ }\textbf {\bibinfo {volume} {368}},\ \bibinfo {pages}
  {489--493} (\bibinfo {year} {2020})}\BibitemShut {NoStop}%
\bibitem [{\citenamefont {Hethcote}(2000)}]{Hethcote2000}%
  \BibitemOpen
  \bibfield  {author} {\bibinfo {author} {\bibfnamefont {H.~W.}\ \bibnamefont
  {Hethcote}},\ }\bibfield  {title} {\enquote {\bibinfo {title} {The
  mathematics of infectious diseases},}\ }\href
  {https://doi.org/10.1137/s0036144500371907} {\bibfield  {journal} {\bibinfo
  {journal} {{SIAM} Review}\ }\textbf {\bibinfo {volume} {42}},\ \bibinfo
  {pages} {599--653} (\bibinfo {year} {2000})}\BibitemShut {NoStop}%
\bibitem [{\citenamefont {Chowell}\ \emph {et~al.}(2016)\citenamefont
  {Chowell}, \citenamefont {Sattenspiel}, \citenamefont {Bansal},\ and\
  \citenamefont {Viboud}}]{Chowell2016}%
  \BibitemOpen
  \bibfield  {author} {\bibinfo {author} {\bibfnamefont {G.}~\bibnamefont
  {Chowell}}, \bibinfo {author} {\bibfnamefont {L.}~\bibnamefont
  {Sattenspiel}}, \bibinfo {author} {\bibfnamefont {S.}~\bibnamefont
  {Bansal}},\ and\ \bibinfo {author} {\bibfnamefont {C.}~\bibnamefont
  {Viboud}},\ }\bibfield  {title} {\enquote {\bibinfo {title} {Mathematical
  models to characterize early epidemic growth: A review},}\ }\href
  {https://doi.org/10.1016/j.plrev.2016.07.005} {\bibfield  {journal} {\bibinfo
   {journal} {Physics of Life Reviews}\ }\textbf {\bibinfo {volume} {18}},\
  \bibinfo {pages} {66--97} (\bibinfo {year} {2016})}\BibitemShut {NoStop}%
\bibitem [{\citenamefont {Biswas}, \citenamefont {Ghosh},\ and\ \citenamefont
  {Sarkar}(2020)}]{Biswas2020}%
  \BibitemOpen
  \bibfield  {author} {\bibinfo {author} {\bibfnamefont {S.~K.}\ \bibnamefont
  {Biswas}}, \bibinfo {author} {\bibfnamefont {U.}~\bibnamefont {Ghosh}},\ and\
  \bibinfo {author} {\bibfnamefont {S.}~\bibnamefont {Sarkar}},\ }\bibfield
  {title} {\enquote {\bibinfo {title} {Mathematical model of {Z}ika virus
  dynamics with vector control and sensitivity analysis},}\ }\href
  {https://doi.org/10.1016/j.idm.2019.12.001} {\bibfield  {journal} {\bibinfo
  {journal} {Infectious Disease Modelling}\ }\textbf {\bibinfo {volume} {5}},\
  \bibinfo {pages} {23--41} (\bibinfo {year} {2020})}\BibitemShut {NoStop}%
\bibitem [{\citenamefont {Morrison}\ and\ \citenamefont
  {Cunha}(2020)}]{Morrison2020}%
  \BibitemOpen
  \bibfield  {author} {\bibinfo {author} {\bibfnamefont {R.~E.}\ \bibnamefont
  {Morrison}}\ and\ \bibinfo {author} {\bibfnamefont {A.}~\bibnamefont
  {Cunha}},\ }\bibfield  {title} {\enquote {\bibinfo {title} {Embedded model
  discrepancy: A case study of {Z}ika modeling},}\ }\href
  {https://doi.org/10.1063/5.0005204} {\bibfield  {journal} {\bibinfo
  {journal} {Chaos: An Interdisciplinary Journal of Nonlinear Science}\
  }\textbf {\bibinfo {volume} {30}},\ \bibinfo {pages} {051103} (\bibinfo
  {year} {2020})}\BibitemShut {NoStop}%
\bibitem [{\citenamefont {Funk}\ \emph {et~al.}(2018)\citenamefont {Funk},
  \citenamefont {Camacho}, \citenamefont {Kucharski}, \citenamefont {Eggo},\
  and\ \citenamefont {Edmunds}}]{Funk2018}%
  \BibitemOpen
  \bibfield  {author} {\bibinfo {author} {\bibfnamefont {S.}~\bibnamefont
  {Funk}}, \bibinfo {author} {\bibfnamefont {A.}~\bibnamefont {Camacho}},
  \bibinfo {author} {\bibfnamefont {A.~J.}\ \bibnamefont {Kucharski}}, \bibinfo
  {author} {\bibfnamefont {R.~M.}\ \bibnamefont {Eggo}},\ and\ \bibinfo
  {author} {\bibfnamefont {W.~J.}\ \bibnamefont {Edmunds}},\ }\bibfield
  {title} {\enquote {\bibinfo {title} {Real-time forecasting of infectious
  disease dynamics with a stochastic semi-mechanistic model},}\ }\href
  {https://doi.org/10.1016/j.epidem.2016.11.003} {\bibfield  {journal}
  {\bibinfo  {journal} {Epidemics}\ }\textbf {\bibinfo {volume} {22}},\
  \bibinfo {pages} {56--61} (\bibinfo {year} {2018})}\BibitemShut {NoStop}%
\bibitem [{\citenamefont {Mhlanga}(2019)}]{Mhlanga2019}%
  \BibitemOpen
  \bibfield  {author} {\bibinfo {author} {\bibfnamefont {A.}~\bibnamefont
  {Mhlanga}},\ }\bibfield  {title} {\enquote {\bibinfo {title} {Dynamical
  analysis and control strategies in modelling ebola virus disease},}\ }\href
  {https://doi.org/10.1186/s13662-019-2392-x} {\bibfield  {journal} {\bibinfo
  {journal} {Advances in Difference Equations}\ }\textbf {\bibinfo {volume}
  {2019}} (\bibinfo {year} {2019}),\ 10.1186/s13662-019-2392-x}\BibitemShut
  {NoStop}%
\bibitem [{\citenamefont {Manchein}\ \emph {et~al.}(2020)\citenamefont
  {Manchein}, \citenamefont {Brugnago}, \citenamefont {da~Silva}, \citenamefont
  {Mendes},\ and\ \citenamefont {Beims}}]{Manchein2020}%
  \BibitemOpen
  \bibfield  {author} {\bibinfo {author} {\bibfnamefont {C.}~\bibnamefont
  {Manchein}}, \bibinfo {author} {\bibfnamefont {E.~L.}\ \bibnamefont
  {Brugnago}}, \bibinfo {author} {\bibfnamefont {R.~M.}\ \bibnamefont
  {da~Silva}}, \bibinfo {author} {\bibfnamefont {C.~F.~O.}\ \bibnamefont
  {Mendes}},\ and\ \bibinfo {author} {\bibfnamefont {M.~W.}\ \bibnamefont
  {Beims}},\ }\bibfield  {title} {\enquote {\bibinfo {title} {Strong
  correlations between power-law growth of {COVID}-19 in four continents and
  the inefficiency of soft quarantine strategies},}\ }\href
  {https://doi.org/10.1063/5.0009454} {\bibfield  {journal} {\bibinfo
  {journal} {Chaos: An Interdisciplinary Journal of Nonlinear Science}\
  }\textbf {\bibinfo {volume} {30}},\ \bibinfo {pages} {041102} (\bibinfo
  {year} {2020})}\BibitemShut {NoStop}%
\bibitem [{\citenamefont {Machado}\ and\ \citenamefont
  {Lopes}(2020)}]{Machado2020}%
  \BibitemOpen
  \bibfield  {author} {\bibinfo {author} {\bibfnamefont {J.~A.~T.}\
  \bibnamefont {Machado}}\ and\ \bibinfo {author} {\bibfnamefont {A.~M.}\
  \bibnamefont {Lopes}},\ }\bibfield  {title} {\enquote {\bibinfo {title} {Rare
  and extreme events: the case of {COVID}-19 pandemic},}\ }\href
  {https://doi.org/10.1007/s11071-020-05680-w} {\bibfield  {journal} {\bibinfo
  {journal} {Nonlinear Dynamics}\ } (\bibinfo {year} {2020}),\
  10.1007/s11071-020-05680-w}\BibitemShut {NoStop}%
\bibitem [{\citenamefont {James}\ and\ \citenamefont
  {Menzies}(2020{\natexlab{b}})}]{chaos1James2020}%
  \BibitemOpen
  \bibfield  {author} {\bibinfo {author} {\bibfnamefont {N.}~\bibnamefont
  {James}}\ and\ \bibinfo {author} {\bibfnamefont {M.}~\bibnamefont
  {Menzies}},\ }\bibfield  {title} {\enquote {\bibinfo {title} {Cluster-based
  dual evolution for multivariate time series: Analyzing {COVID}-19},}\ }\href
  {https://doi.org/10.1063/5.0013156} {\bibfield  {journal} {\bibinfo
  {journal} {Chaos: An Interdisciplinary Journal of Nonlinear Science}\
  }\textbf {\bibinfo {volume} {30}},\ \bibinfo {pages} {061108} (\bibinfo
  {year} {2020}{\natexlab{b}})}\BibitemShut {NoStop}%
\bibitem [{\citenamefont {Blasius}(2020)}]{Blasius2020}%
  \BibitemOpen
  \bibfield  {author} {\bibinfo {author} {\bibfnamefont {B.}~\bibnamefont
  {Blasius}},\ }\bibfield  {title} {\enquote {\bibinfo {title} {Power-law
  distribution in the number of confirmed {COVID}-19 cases},}\ }\href
  {https://doi.org/10.1063/5.0013031} {\bibfield  {journal} {\bibinfo
  {journal} {Chaos: An Interdisciplinary Journal of Nonlinear Science}\
  }\textbf {\bibinfo {volume} {30}},\ \bibinfo {pages} {093123} (\bibinfo
  {year} {2020})}\BibitemShut {NoStop}%
\bibitem [{\citenamefont {Perc}\ \emph {et~al.}(2020)\citenamefont {Perc},
  \citenamefont {Miksi{\'{c}}}, \citenamefont {Slavinec},\ and\ \citenamefont
  {Sto{\v{z}}er}}]{Perc2020}%
  \BibitemOpen
  \bibfield  {author} {\bibinfo {author} {\bibfnamefont {M.}~\bibnamefont
  {Perc}}, \bibinfo {author} {\bibfnamefont {N.~G.}\ \bibnamefont
  {Miksi{\'{c}}}}, \bibinfo {author} {\bibfnamefont {M.}~\bibnamefont
  {Slavinec}},\ and\ \bibinfo {author} {\bibfnamefont {A.}~\bibnamefont
  {Sto{\v{z}}er}},\ }\bibfield  {title} {\enquote {\bibinfo {title}
  {Forecasting {COVID}-19},}\ }\href {https://doi.org/10.3389/fphy.2020.00127}
  {\bibfield  {journal} {\bibinfo  {journal} {Frontiers in Physics}\ }\textbf
  {\bibinfo {volume} {8}},\ \bibinfo {pages} {127} (\bibinfo {year}
  {2020})}\BibitemShut {NoStop}%
\bibitem [{\citenamefont {Vazquez}(2006)}]{Vazquez2006}%
  \BibitemOpen
  \bibfield  {author} {\bibinfo {author} {\bibfnamefont {A.}~\bibnamefont
  {Vazquez}},\ }\bibfield  {title} {\enquote {\bibinfo {title} {Polynomial
  growth in branching processes with diverging reproductive number},}\ }\href
  {https://doi.org/10.1103/physrevlett.96.038702} {\bibfield  {journal}
  {\bibinfo  {journal} {Physical Review Letters}\ }\textbf {\bibinfo {volume}
  {96}} (\bibinfo {year} {2006}),\ 10.1103/physrevlett.96.038702}\BibitemShut
  {NoStop}%
\bibitem [{\citenamefont {Moeckel}\ and\ \citenamefont
  {Murray}(1997)}]{Moeckel1997}%
  \BibitemOpen
  \bibfield  {author} {\bibinfo {author} {\bibfnamefont {R.}~\bibnamefont
  {Moeckel}}\ and\ \bibinfo {author} {\bibfnamefont {B.}~\bibnamefont
  {Murray}},\ }\bibfield  {title} {\enquote {\bibinfo {title} {Measuring the
  distance between time series},}\ }\href
  {https://doi.org/10.1016/s0167-2789(96)00154-6} {\bibfield  {journal}
  {\bibinfo  {journal} {Physica D: Nonlinear Phenomena}\ }\textbf {\bibinfo
  {volume} {102}},\ \bibinfo {pages} {187--194} (\bibinfo {year}
  {1997})}\BibitemShut {NoStop}%
\bibitem [{\citenamefont {Sz{\'{e}}kely}, \citenamefont {Rizzo},\ and\
  \citenamefont {Bakirov}(2007)}]{Szkely2007}%
  \BibitemOpen
  \bibfield  {author} {\bibinfo {author} {\bibfnamefont {G.~J.}\ \bibnamefont
  {Sz{\'{e}}kely}}, \bibinfo {author} {\bibfnamefont {M.~L.}\ \bibnamefont
  {Rizzo}},\ and\ \bibinfo {author} {\bibfnamefont {N.~K.}\ \bibnamefont
  {Bakirov}},\ }\bibfield  {title} {\enquote {\bibinfo {title} {Measuring and
  testing dependence by correlation of distances},}\ }\href
  {https://doi.org/10.1214/009053607000000505} {\bibfield  {journal} {\bibinfo
  {journal} {The Annals of Statistics}\ }\textbf {\bibinfo {volume} {35}},\
  \bibinfo {pages} {2769--2794} (\bibinfo {year} {2007})}\BibitemShut {NoStop}%
\bibitem [{\citenamefont {Mendes}\ and\ \citenamefont
  {Beims}(2018)}]{Mendes2018}%
  \BibitemOpen
  \bibfield  {author} {\bibinfo {author} {\bibfnamefont {C.~F.}\ \bibnamefont
  {Mendes}}\ and\ \bibinfo {author} {\bibfnamefont {M.~W.}\ \bibnamefont
  {Beims}},\ }\bibfield  {title} {\enquote {\bibinfo {title} {Distance
  correlation detecting {L}yapunov instabilities, noise-induced escape times
  and mixing},}\ }\href {https://doi.org/10.1016/j.physa.2018.08.028}
  {\bibfield  {journal} {\bibinfo  {journal} {Physica A: Statistical Mechanics
  and its Applications}\ }\textbf {\bibinfo {volume} {512}},\ \bibinfo {pages}
  {721--730} (\bibinfo {year} {2018})}\BibitemShut {NoStop}%
\bibitem [{\citenamefont {Mendes}, \citenamefont {da~Silva},\ and\
  \citenamefont {Beims}(2019)}]{Mendes2019}%
  \BibitemOpen
  \bibfield  {author} {\bibinfo {author} {\bibfnamefont {C.~F.~O.}\
  \bibnamefont {Mendes}}, \bibinfo {author} {\bibfnamefont {R.~M.}\
  \bibnamefont {da~Silva}},\ and\ \bibinfo {author} {\bibfnamefont {M.~W.}\
  \bibnamefont {Beims}},\ }\bibfield  {title} {\enquote {\bibinfo {title}
  {Decay of the distance autocorrelation and {L}yapunov exponents},}\ }\href
  {https://doi.org/10.1103/physreve.99.062206} {\bibfield  {journal} {\bibinfo
  {journal} {Physical Review E}\ }\textbf {\bibinfo {volume} {99}} (\bibinfo
  {year} {2019}),\ 10.1103/physreve.99.062206}\BibitemShut {NoStop}%
\bibitem [{\citenamefont {James}\ \emph {et~al.}(2020)\citenamefont {James},
  \citenamefont {Menzies}, \citenamefont {Azizi},\ and\ \citenamefont
  {Chan}}]{James2020_nsm}%
  \BibitemOpen
  \bibfield  {author} {\bibinfo {author} {\bibfnamefont {N.}~\bibnamefont
  {James}}, \bibinfo {author} {\bibfnamefont {M.}~\bibnamefont {Menzies}},
  \bibinfo {author} {\bibfnamefont {L.}~\bibnamefont {Azizi}},\ and\ \bibinfo
  {author} {\bibfnamefont {J.}~\bibnamefont {Chan}},\ }\bibfield  {title}
  {\enquote {\bibinfo {title} {Novel semi-metrics for multivariate change point
  analysis and anomaly detection},}\ }\href
  {https://doi.org/10.1016/j.physd.2020.132636} {\bibfield  {journal} {\bibinfo
   {journal} {Physica D: Nonlinear Phenomena}\ }\textbf {\bibinfo {volume}
  {412}},\ \bibinfo {pages} {132636} (\bibinfo {year} {2020})}\BibitemShut
  {NoStop}%
\bibitem [{\citenamefont {Shang}\ \emph {et~al.}(2020)\citenamefont {Shang},
  \citenamefont {Yang}, \citenamefont {Moore}, \citenamefont {Ji},\ and\
  \citenamefont {Small}}]{Shang2020}%
  \BibitemOpen
  \bibfield  {author} {\bibinfo {author} {\bibfnamefont {K.}~\bibnamefont
  {Shang}}, \bibinfo {author} {\bibfnamefont {B.}~\bibnamefont {Yang}},
  \bibinfo {author} {\bibfnamefont {J.~M.}\ \bibnamefont {Moore}}, \bibinfo
  {author} {\bibfnamefont {Q.}~\bibnamefont {Ji}},\ and\ \bibinfo {author}
  {\bibfnamefont {M.}~\bibnamefont {Small}},\ }\bibfield  {title} {\enquote
  {\bibinfo {title} {Growing networks with communities: A distributive link
  model},}\ }\href {https://doi.org/10.1063/5.0007422} {\bibfield  {journal}
  {\bibinfo  {journal} {Chaos: An Interdisciplinary Journal of Nonlinear
  Science}\ }\textbf {\bibinfo {volume} {30}},\ \bibinfo {pages} {041101}
  (\bibinfo {year} {2020})}\BibitemShut {NoStop}%
\bibitem [{\citenamefont {Karaivanov}(2020)}]{Karaivanov2020}%
  \BibitemOpen
  \bibfield  {author} {\bibinfo {author} {\bibfnamefont {A.}~\bibnamefont
  {Karaivanov}},\ }\bibfield  {title} {\enquote {\bibinfo {title} {A social
  network model of {COVID}-19},}\ }\href
  {https://doi.org/10.1371/journal.pone.0240878} {\bibfield  {journal}
  {\bibinfo  {journal} {{PLOS} {ONE}}\ }\textbf {\bibinfo {volume} {15}},\
  \bibinfo {pages} {e0240878} (\bibinfo {year} {2020})}\BibitemShut {NoStop}%
\bibitem [{\citenamefont {Madore}\ \emph {et~al.}(2007)\citenamefont {Madore}
  \emph {et~al.}}]{Madore2007}%
  \BibitemOpen
  \bibfield  {author} {\bibinfo {author} {\bibfnamefont {A.-M.}\ \bibnamefont
  {Madore}} \emph {et~al.},\ }\bibfield  {title} {\enquote {\bibinfo {title}
  {Contribution of hierarchical clustering techniques to the modeling of the
  geographic distribution of genetic polymorphisms associated with chronic
  inflammatory diseases in the {Q}u{\'{e}}bec population},}\ }\href
  {https://doi.org/10.1159/000106560} {\bibfield  {journal} {\bibinfo
  {journal} {Public Health Genomics}\ }\textbf {\bibinfo {volume} {10}},\
  \bibinfo {pages} {218--226} (\bibinfo {year} {2007})}\BibitemShut {NoStop}%
\bibitem [{\citenamefont {Kretzschmar}\ and\ \citenamefont
  {Mikolajczyk}(2009)}]{Kretzschmar2009}%
  \BibitemOpen
  \bibfield  {author} {\bibinfo {author} {\bibfnamefont {M.}~\bibnamefont
  {Kretzschmar}}\ and\ \bibinfo {author} {\bibfnamefont {R.~T.}\ \bibnamefont
  {Mikolajczyk}},\ }\bibfield  {title} {\enquote {\bibinfo {title} {Contact
  profiles in eight {E}uropean countries and implications for modelling the
  spread of airborne infectious diseases},}\ }\href
  {https://doi.org/10.1371/journal.pone.0005931} {\bibfield  {journal}
  {\bibinfo  {journal} {{PLoS} {ONE}}\ }\textbf {\bibinfo {volume} {4}},\
  \bibinfo {pages} {e5931} (\bibinfo {year} {2009})}\BibitemShut {NoStop}%
\bibitem [{\citenamefont {Alashwal}\ \emph {et~al.}(2019)\citenamefont
  {Alashwal}, \citenamefont {Halaby}, \citenamefont {Crouse}, \citenamefont
  {Abdalla},\ and\ \citenamefont {Moustafa}}]{Alashwal2019}%
  \BibitemOpen
  \bibfield  {author} {\bibinfo {author} {\bibfnamefont {H.}~\bibnamefont
  {Alashwal}}, \bibinfo {author} {\bibfnamefont {M.~E.}\ \bibnamefont
  {Halaby}}, \bibinfo {author} {\bibfnamefont {J.~J.}\ \bibnamefont {Crouse}},
  \bibinfo {author} {\bibfnamefont {A.}~\bibnamefont {Abdalla}},\ and\ \bibinfo
  {author} {\bibfnamefont {A.~A.}\ \bibnamefont {Moustafa}},\ }\bibfield
  {title} {\enquote {\bibinfo {title} {The application of unsupervised
  clustering methods to {A}lzheimer's disease},}\ }\href
  {https://doi.org/10.3389/fncom.2019.00031} {\bibfield  {journal} {\bibinfo
  {journal} {Frontiers in Computational Neuroscience}\ }\textbf {\bibinfo
  {volume} {13}} (\bibinfo {year} {2019}),\
  10.3389/fncom.2019.00031}\BibitemShut {NoStop}%
\bibitem [{\citenamefont {Muradi}, \citenamefont {Bustamam},\ and\
  \citenamefont {Lestari}(2015)}]{Muradi2015}%
  \BibitemOpen
  \bibfield  {author} {\bibinfo {author} {\bibfnamefont {H.}~\bibnamefont
  {Muradi}}, \bibinfo {author} {\bibfnamefont {A.}~\bibnamefont {Bustamam}},\
  and\ \bibinfo {author} {\bibfnamefont {D.}~\bibnamefont {Lestari}},\
  }\bibfield  {title} {\enquote {\bibinfo {title} {Application of hierarchical
  clustering ordered partitioning and collapsing hybrid in {E}bola virus
  phylogenetic analysis},}\ }in\ \href
  {https://doi.org/10.1109/icacsis.2015.7415183} {\emph {\bibinfo {booktitle}
  {2015 International Conference on Advanced Computer Science and Information
  Systems ({ICACSIS})}}}\ (\bibinfo  {publisher} {{IEEE}},\ \bibinfo {year}
  {2015})\BibitemShut {NoStop}%
\bibitem [{\citenamefont {Rizzi}\ \emph {et~al.}(2010)\citenamefont {Rizzi},
  \citenamefont {Mahata}, \citenamefont {Mathieson},\ and\ \citenamefont
  {Moscato}}]{Rizzi2010}%
  \BibitemOpen
  \bibfield  {author} {\bibinfo {author} {\bibfnamefont {R.}~\bibnamefont
  {Rizzi}}, \bibinfo {author} {\bibfnamefont {P.}~\bibnamefont {Mahata}},
  \bibinfo {author} {\bibfnamefont {L.}~\bibnamefont {Mathieson}},\ and\
  \bibinfo {author} {\bibfnamefont {P.}~\bibnamefont {Moscato}},\ }\bibfield
  {title} {\enquote {\bibinfo {title} {Hierarchical clustering using the
  arithmetic-harmonic cut: Complexity and experiments},}\ }\href
  {https://doi.org/10.1371/journal.pone.0014067} {\bibfield  {journal}
  {\bibinfo  {journal} {{PLoS} {ONE}}\ }\textbf {\bibinfo {volume} {5}},\
  \bibinfo {pages} {e14067} (\bibinfo {year} {2010})}\BibitemShut {NoStop}%
\bibitem [{Eur(2020)}]{Europelist}%
  \BibitemOpen
  \href@noop {} {\enquote {\bibinfo {title} {Worldodometer},}\ }\bibinfo
  {howpublished}
  {\url{https://www.worldometers.info/geography/how-many-countries-in-europe/}}
  (\bibinfo {year} {2020}),\ \bibinfo {note} {accessed November 25,
  2020}\BibitemShut {NoStop}%
\bibitem [{dat(2020)}]{datachanged}%
  \BibitemOpen
  \href@noop {} {\enquote {\bibinfo {title} {Our {W}orld in {D}ata},}\
  }\bibinfo {howpublished}
  {\url{https://ourworldindata.org/covid-data-switch-jhu}} (\bibinfo {year}
  {2020}),\ \bibinfo {note} {accessed November 25, 2020}\BibitemShut {NoStop}%
\bibitem [{\citenamefont {Avis}, \citenamefont {Bremner},\ and\ \citenamefont
  {Seidel}(1997)}]{convexhull}%
  \BibitemOpen
  \bibfield  {author} {\bibinfo {author} {\bibfnamefont {D.}~\bibnamefont
  {Avis}}, \bibinfo {author} {\bibfnamefont {D.}~\bibnamefont {Bremner}},\ and\
  \bibinfo {author} {\bibfnamefont {R.}~\bibnamefont {Seidel}},\ }\bibfield
  {title} {\enquote {\bibinfo {title} {How good are convex hull algorithms?}}\
  }\href {https://doi.org/10.1016/s0925-7721(96)00023-5} {\bibfield  {journal}
  {\bibinfo  {journal} {Computational Geometry}\ }\textbf {\bibinfo {volume}
  {7}},\ \bibinfo {pages} {265--301} (\bibinfo {year} {1997})}\BibitemShut
  {NoStop}%
\bibitem [{\citenamefont {Campana}()}]{Greecetourism}%
  \BibitemOpen
  \bibfield  {author} {\bibinfo {author} {\bibfnamefont {F.}~\bibnamefont
  {Campana}},\ }\href@noop {} {\enquote {\bibinfo {title} {How greece is
  rethinking its once bustling tourism industry},}\ }\bibinfo {howpublished}
  {\url{https://www.nationalgeographic.com/travel/2020/09/how-greece-is-coping-without-tourism-due-to-covid/
  }},\ \bibinfo {note} {{National Geographic}, September 22, 2020}\BibitemShut
  {NoStop}%
\bibitem [{\citenamefont {Kontopantelis}\ \emph {et~al.}(2020)\citenamefont
  {Kontopantelis}, \citenamefont {Mamas}, \citenamefont {Deanfield},
  \citenamefont {Asaria},\ and\ \citenamefont {Doran}}]{Kontopantelis2020}%
  \BibitemOpen
  \bibfield  {author} {\bibinfo {author} {\bibfnamefont {E.}~\bibnamefont
  {Kontopantelis}}, \bibinfo {author} {\bibfnamefont {M.~A.}\ \bibnamefont
  {Mamas}}, \bibinfo {author} {\bibfnamefont {J.}~\bibnamefont {Deanfield}},
  \bibinfo {author} {\bibfnamefont {M.}~\bibnamefont {Asaria}},\ and\ \bibinfo
  {author} {\bibfnamefont {T.}~\bibnamefont {Doran}},\ }\bibfield  {title}
  {\enquote {\bibinfo {title} {Excess mortality in {E}ngland and {W}ales during
  the first wave of the {COVID}-19 pandemic},}\ }\href
  {https://doi.org/10.1136/jech-2020-214764} {\bibfield  {journal} {\bibinfo
  {journal} {Journal of Epidemiology and Community Health}\ ,\ \bibinfo {pages}
  {jech--2020--214764}} (\bibinfo {year} {2020})}\BibitemShut {NoStop}%
\bibitem [{\citenamefont {Aleta}\ and\ \citenamefont
  {Moreno}(2020)}]{Aleta2020}%
  \BibitemOpen
  \bibfield  {author} {\bibinfo {author} {\bibfnamefont {A.}~\bibnamefont
  {Aleta}}\ and\ \bibinfo {author} {\bibfnamefont {Y.}~\bibnamefont {Moreno}},\
  }\bibfield  {title} {\enquote {\bibinfo {title} {Age differential analysis of
  {COVID}-19 second wave in {E}urope reveals highest incidence among young
  adults},}\ }\href {https://doi.org/10.1101/2020.11.11.20230177} {\bibfield
  {journal} {\bibinfo  {journal} {medRxiv}\ } (\bibinfo {year} {2020}),\
  10.1101/2020.11.11.20230177}\BibitemShut {NoStop}%
\bibitem [{\citenamefont {Burke}\ and\ \citenamefont
  {Ryan}(2014)}]{UShealthcare}%
  \BibitemOpen
  \bibfield  {author} {\bibinfo {author} {\bibfnamefont {L.~A.}\ \bibnamefont
  {Burke}}\ and\ \bibinfo {author} {\bibfnamefont {A.~M.}\ \bibnamefont
  {Ryan}},\ }\bibfield  {title} {\enquote {\bibinfo {title} {The complex
  relationship between cost and quality in {US} health care},}\ }\href
  {https://doi.org/10.1001/virtualmentor.2014.16.2.pfor1-1402} {\bibfield
  {journal} {\bibinfo  {journal} {{AMA} Journal of Ethics}\ }\textbf {\bibinfo
  {volume} {16}},\ \bibinfo {pages} {124--130} (\bibinfo {year}
  {2014})}\BibitemShut {NoStop}%
\bibitem [{Fra()}]{Francechange}%
  \BibitemOpen
  \href@noop {} {\enquote {\bibinfo {title} {Prise en charge par les
  m{\'e}dicins de ville des patients atteints de {COVID}-19 en phase de
  d{\'e}confinement},}\ }\bibinfo {howpublished}
  {\url{https://solidarites-sante.gouv.fr/IMG/pdf/prise-en-charge-medecine-ville-covid-19.pdf}},\
  \bibinfo {note} {{M}inist{\`e}re des {S}olidarit{\'e}s et de la {S}ant{\'e},
  May 13, 2020}\BibitemShut {NoStop}%
\bibitem [{\citenamefont {Pullano}\ \emph {et~al.}(2020)\citenamefont {Pullano}
  \emph {et~al.}}]{Pullano2020}%
  \BibitemOpen
  \bibfield  {author} {\bibinfo {author} {\bibfnamefont {G.}~\bibnamefont
  {Pullano}} \emph {et~al.},\ }\bibfield  {title} {\enquote {\bibinfo {title}
  {Underdetection of cases of {COVID}-19 in {F}rance threatens epidemic
  control},}\ }\href {https://doi.org/10.1038/s41586-020-03095-6} {\bibfield
  {journal} {\bibinfo  {journal} {Nature}\ } (\bibinfo {year} {2020}),\
  10.1038/s41586-020-03095-6}\BibitemShut {NoStop}%
\bibitem [{\citenamefont {Miller}, \citenamefont {Copley},\ and\ \citenamefont
  {Meijer}()}]{Antigenchange}%
  \BibitemOpen
  \bibfield  {author} {\bibinfo {author} {\bibfnamefont {J.}~\bibnamefont
  {Miller}}, \bibinfo {author} {\bibfnamefont {C.}~\bibnamefont {Copley}},\
  and\ \bibinfo {author} {\bibfnamefont {B.~H.}\ \bibnamefont {Meijer}},\
  }\href@noop {} {\enquote {\bibinfo {title} {Countries turn to rapid antigen
  tests to contain second wave of {COVID}-19},}\ }\bibinfo {howpublished}
  {\url{https://www.reuters.com/article/idUSKBN26Z2C2}},\ \bibinfo {note}
  {{R}euters, October 15, 2020}\BibitemShut {NoStop}%
\bibitem [{\citenamefont {Bari}\ \emph {et~al.}(2020)\citenamefont {Bari},
  \citenamefont {Balzi}, \citenamefont {Carreras},\ and\ \citenamefont
  {Onder}}]{DiBari2020}%
  \BibitemOpen
  \bibfield  {author} {\bibinfo {author} {\bibfnamefont {M.~D.}\ \bibnamefont
  {Bari}}, \bibinfo {author} {\bibfnamefont {D.}~\bibnamefont {Balzi}},
  \bibinfo {author} {\bibfnamefont {G.}~\bibnamefont {Carreras}},\ and\
  \bibinfo {author} {\bibfnamefont {G.}~\bibnamefont {Onder}},\ }\bibfield
  {title} {\enquote {\bibinfo {title} {Extensive testing may reduce {COVID}-19
  mortality: A lesson from northern {I}taly},}\ }\href
  {https://doi.org/10.3389/fmed.2020.00402} {\bibfield  {journal} {\bibinfo
  {journal} {Frontiers in Medicine}\ }\textbf {\bibinfo {volume} {7}} (\bibinfo
  {year} {2020}),\ 10.3389/fmed.2020.00402}\BibitemShut {NoStop}%
\bibitem [{ita()}]{italydeaths}%
  \BibitemOpen
  \href@noop {} {\enquote {\bibinfo {title} {Impatto dell’epidemia {COVID}-19
  sulla mortalit{\`a} totale della popolazione residente primo quadrimestre
  2020},}\ }\bibinfo {howpublished}
  {\url{https://www.epicentro.iss.it/coronavirus/pdf/Rapp_Istat_Iss_3Giugno.pdf}},\
  \bibinfo {note} {{Instituto Nazionale di Statistica}, June 4,
  2020}\BibitemShut {NoStop}%
\bibitem [{\citenamefont {Reiner}\ \emph {et~al.}(2021)\citenamefont {Reiner}
  \emph {et~al.}}]{IHMEmodel}%
  \BibitemOpen
  \bibfield  {author} {\bibinfo {author} {\bibfnamefont {R.~C.}\ \bibnamefont
  {Reiner}} \emph {et~al.},\ }\bibfield  {title} {\enquote {\bibinfo {title}
  {Modeling {COVID}-19 scenarios for the {U}nited {S}tates},}\ }\href
  {https://doi.org/10.1038/s41591-020-1132-9} {\bibfield  {journal} {\bibinfo
  {journal} {Nature Medicine}\ }\textbf {\bibinfo {volume} {27}},\ \bibinfo
  {pages} {94--105} (\bibinfo {year} {2021})}\BibitemShut {NoStop}%
\bibitem [{\citenamefont {Momtazmanesh}\ \emph {et~al.}(2020)\citenamefont
  {Momtazmanesh} \emph {et~al.}}]{Momtazmanesh2020}%
  \BibitemOpen
  \bibfield  {author} {\bibinfo {author} {\bibfnamefont {S.}~\bibnamefont
  {Momtazmanesh}} \emph {et~al.},\ }\bibfield  {title} {\enquote {\bibinfo
  {title} {All together to fight {COVID}-19},}\ }\href
  {https://doi.org/10.4269/ajtmh.20-0281} {\bibfield  {journal} {\bibinfo
  {journal} {The American Journal of Tropical Medicine and Hygiene}\ }\textbf
  {\bibinfo {volume} {102}},\ \bibinfo {pages} {1181--1183} (\bibinfo {year}
  {2020})}\BibitemShut {NoStop}%
\bibitem [{\citenamefont {Priesemann}\ \emph {et~al.}(2021)\citenamefont
  {Priesemann} \emph {et~al.}}]{Priesemann2021}%
  \BibitemOpen
  \bibfield  {author} {\bibinfo {author} {\bibfnamefont {V.}~\bibnamefont
  {Priesemann}} \emph {et~al.},\ }\bibfield  {title} {\enquote {\bibinfo
  {title} {Calling for pan-{E}uropean commitment for rapid and sustained
  reduction in {SARS}-{CoV}-2 infections},}\ }\href
  {https://doi.org/10.1016/s0140-6736(20)32625-8} {\bibfield  {journal}
  {\bibinfo  {journal} {The Lancet}\ }\textbf {\bibinfo {volume} {397}},\
  \bibinfo {pages} {92--93} (\bibinfo {year} {2021})}\BibitemShut {NoStop}%
\bibitem [{wor(2020)}]{worldindata2020}%
  \BibitemOpen
  \href@noop {} {\enquote {\bibinfo {title} {Our {W}orld in {D}ata},}\
  }\bibinfo {howpublished}
  {\url{https://ourworldindata.org/coronavirus-source-data}} (\bibinfo {year}
  {2020}),\ \bibinfo {note} {accessed November 25, 2020}\BibitemShut {NoStop}%
\bibitem [{dat()}]{datasource}%
  \BibitemOpen
  \href@noop {} {\enquote {\bibinfo {title} {Coronavirus {(Covid-19)} data in
  the {U}nited {S}tates},}\ }\bibinfo {howpublished}
  {\url{https://github.com/nytimes/covid-19-data}},\ \bibinfo {note} {{The New
  York Times}, Accessed November 25, 2020}\BibitemShut {NoStop}%
\bibitem [{\citenamefont {Lauer}\ \emph {et~al.}(2020)\citenamefont {Lauer}
  \emph {et~al.}}]{incubation2020}%
  \BibitemOpen
  \bibfield  {author} {\bibinfo {author} {\bibfnamefont {S.~A.}\ \bibnamefont
  {Lauer}} \emph {et~al.},\ }\bibfield  {title} {\enquote {\bibinfo {title}
  {The incubation period of {C}oronavirus disease 2019 ({COVID}-19) from
  publicly reported confirmed cases: Estimation and application},}\ }\href
  {https://doi.org/10.7326/m20-0504} {\bibfield  {journal} {\bibinfo  {journal}
  {Annals of Internal Medicine}\ }\textbf {\bibinfo {volume} {172}},\ \bibinfo
  {pages} {577--582} (\bibinfo {year} {2020})}\BibitemShut {NoStop}%
\bibitem [{\citenamefont {Blamont}()}]{fludeaths}%
  \BibitemOpen
  \bibfield  {author} {\bibinfo {author} {\bibfnamefont {M.}~\bibnamefont
  {Blamont}},\ }\href@noop {} {\enquote {\bibinfo {title} {Analysis: Could
  {COVID} knock out flu in {E}urope this winter?}}\ }\bibinfo {howpublished}
  {\url{https://www.reuters.com/article/idUKKBN28B531}},\ \bibinfo {note}
  {{R}euters, December 2, 2020}\BibitemShut {NoStop}%
\bibitem [{\citenamefont {Qureshi}\ \emph {et~al.}(2020)\citenamefont
  {Qureshi}, \citenamefont {Huang}, \citenamefont {Khan}, \citenamefont
  {Lobanova}, \citenamefont {Siddiq}, \citenamefont {Gomez},\ and\
  \citenamefont {Suri}}]{Qureshi2020}%
  \BibitemOpen
  \bibfield  {author} {\bibinfo {author} {\bibfnamefont {A.~I.}\ \bibnamefont
  {Qureshi}}, \bibinfo {author} {\bibfnamefont {W.}~\bibnamefont {Huang}},
  \bibinfo {author} {\bibfnamefont {S.}~\bibnamefont {Khan}}, \bibinfo {author}
  {\bibfnamefont {I.}~\bibnamefont {Lobanova}}, \bibinfo {author}
  {\bibfnamefont {F.}~\bibnamefont {Siddiq}}, \bibinfo {author} {\bibfnamefont
  {C.~R.}\ \bibnamefont {Gomez}},\ and\ \bibinfo {author} {\bibfnamefont
  {M.~F.~K.}\ \bibnamefont {Suri}},\ }\bibfield  {title} {\enquote {\bibinfo
  {title} {Mandated societal lockdown and road traffic accidents},}\ }\href
  {https://doi.org/10.1016/j.aap.2020.105747} {\bibfield  {journal} {\bibinfo
  {journal} {Accident Analysis {\&} Prevention}\ }\textbf {\bibinfo {volume}
  {146}},\ \bibinfo {pages} {105747} (\bibinfo {year} {2020})}\BibitemShut
  {NoStop}%
\bibitem [{\citenamefont {Saladi{\'{e}}}, \citenamefont {Bustamante},\ and\
  \citenamefont {Guti{\'{e}}rrez}(2020)}]{Saladi2020}%
  \BibitemOpen
  \bibfield  {author} {\bibinfo {author} {\bibfnamefont {{\`{O}}.}~\bibnamefont
  {Saladi{\'{e}}}}, \bibinfo {author} {\bibfnamefont {E.}~\bibnamefont
  {Bustamante}},\ and\ \bibinfo {author} {\bibfnamefont {A.}~\bibnamefont
  {Guti{\'{e}}rrez}},\ }\bibfield  {title} {\enquote {\bibinfo {title}
  {{COVID}-19 lockdown and reduction of traffic accidents in {T}arragona
  province, {S}pain},}\ }\href {https://doi.org/10.1016/j.trip.2020.100218}
  {\bibfield  {journal} {\bibinfo  {journal} {Transportation Research
  Interdisciplinary Perspectives}\ }\textbf {\bibinfo {volume} {8}},\ \bibinfo
  {pages} {100218} (\bibinfo {year} {2020})}\BibitemShut {NoStop}%
\bibitem [{\citenamefont {John}\ \emph {et~al.}(2020)\citenamefont {John},
  \citenamefont {Pirkis}, \citenamefont {Gunnell}, \citenamefont {Appleby},\
  and\ \citenamefont {Morrissey}}]{John2020}%
  \BibitemOpen
  \bibfield  {author} {\bibinfo {author} {\bibfnamefont {A.}~\bibnamefont
  {John}}, \bibinfo {author} {\bibfnamefont {J.}~\bibnamefont {Pirkis}},
  \bibinfo {author} {\bibfnamefont {D.}~\bibnamefont {Gunnell}}, \bibinfo
  {author} {\bibfnamefont {L.}~\bibnamefont {Appleby}},\ and\ \bibinfo {author}
  {\bibfnamefont {J.}~\bibnamefont {Morrissey}},\ }\bibfield  {title} {\enquote
  {\bibinfo {title} {Trends in suicide during the {C}ovid-19 pandemic},}\
  }\href {https://doi.org/10.1136/bmj.m4352} {\bibfield  {journal} {\bibinfo
  {journal} {{BMJ}}\ ,\ \bibinfo {pages} {m4352}} (\bibinfo {year}
  {2020})}\BibitemShut {NoStop}%
\bibitem [{\citenamefont {Le}\ \emph {et~al.}(2020)\citenamefont {Le},
  \citenamefont {Khan}, \citenamefont {Murtaza},\ and\ \citenamefont
  {Shah}}]{Le2020}%
  \BibitemOpen
  \bibfield  {author} {\bibinfo {author} {\bibfnamefont {H.}~\bibnamefont
  {Le}}, \bibinfo {author} {\bibfnamefont {B.~A.}\ \bibnamefont {Khan}},
  \bibinfo {author} {\bibfnamefont {S.}~\bibnamefont {Murtaza}},\ and\ \bibinfo
  {author} {\bibfnamefont {A.~A.}\ \bibnamefont {Shah}},\ }\bibfield  {title}
  {\enquote {\bibinfo {title} {The increase in suicide during the {COVID}-19
  pandemic},}\ }\href {https://doi.org/10.3928/00485713-20201105-01} {\bibfield
   {journal} {\bibinfo  {journal} {Psychiatric Annals}\ }\textbf {\bibinfo
  {volume} {50}},\ \bibinfo {pages} {526--530} (\bibinfo {year}
  {2020})}\BibitemShut {NoStop}%
\end{thebibliography}%

\end{document}